\documentclass[12pt,a4paper, amsmath,amssymb]{article}
\usepackage[margin=0.75in]{geometry}
\usepackage{longtable}
\usepackage{amssymb}
\usepackage{amsmath}
\usepackage{amsfonts}
\usepackage{bm}
\usepackage[utf8]{inputenc}
\usepackage{graphicx}
\usepackage{float}
\usepackage{xcolor}
\definecolor{bluemunsell}{rgb}{0.0, 0.5, 0.69}
\usepackage[colorlinks,linkcolor = bluemunsell,
urlcolor  = bluemunsell,
citecolor = bluemunsell,
anchorcolor = bluemunsell]{hyperref}
\usepackage{longtable}
\usepackage[normalem]{ulem}

\usepackage{xcolor}
\usepackage[normalem]{ulem}

\usepackage{MnSymbol,bbding,pifont}
\usepackage{xcolor}
\usepackage{makecell}
\usepackage{ctable} 
\usepackage{adjustbox}
\def\thefootnote{\fnsymbol{footnote}}

\usepackage{cite}
\definecolor{mediumjunglegreen}{rgb}{0.11, 0.21, 0.18}
\definecolor{green}{rgb}{0.0, 0.5, 0.0}

\begin{document}
\begin{center}
{\Large \textbf{
Investigating the 95 GeV Higgs Boson Excesses \\ \vspace{0.15cm} within the I(1+2)HDM}}\\
\thispagestyle{empty} 
\vspace{1cm}
{\sc
Ayoub Hmissou $^{1}$\footnote{\href{mailto:ayoub1hmissou@gmail.com}{ayoub1hmissou@gmail.com}},
Stefano Moretti $^{2,3}$\footnote{\href{mailto:stefano.moretti@physics.uu.se/s.moretti@soton.ac.uk}{stefano.moretti@physics.uu.se/s.moretti@soton.ac.uk}},
Larbi Rahili $^1$\footnote{\href{mailto:l.rahili@uiz.ac.ma/rahililarbi@gmail.com}{l.rahili@uiz.ac.ma/rahililarbi@gmail.com}}
\\
}
\vspace{1cm}
{\sl
$^1$Laboratory of Theoretical and High Energy Physics,Faculty of Science, Ibnou Zohr University, B.P 8106, Agadir, Morocco.\\
$^2$Department of Physics \& Astronomy, Uppsala University, Box 516, SE-751 20 Uppsala, Sweden.\\
$^3$School of Physics \& Astronomy, University of Southampton, Southampton, SO17 1BJ, United Kingdom.\\
}
\end{center}
\vspace*{0.1cm}
\begin{abstract}
In this work, we explore how the 2-Higgs Doublet Model (2HDM) Type-I, extended by an inert doublet, can provide an explanation for the recently observed excesses at the Large Hadron Collider (LHC) in the $\gamma\gamma$ and $\tau^+ \tau^- $ final states. 
Hence, by imposing theoretical constraints and experimental bounds on the model parameter space, our findings show that a light CP-even Higgs boson, $h$, with a mass around 95 GeV, can account for these anomalies. This result aligns with the excess in $b\bar b$ signatures reported in earlier data from the Large Electron-Positron (LEP) collider. 
\end{abstract}

\def\thefootnote{\arabic{footnote}}
\setcounter{page}{0}
\setcounter{footnote}{0}

\section{Introduction}
\label{sec:intro}
Since the discovery of a Higgs boson at the LHC in July 2012 \cite{ATLAS:2012yve,CMS:2012qbp}, several analyses of experimental data have reported intriguing anomalies that seem to deviate slightly from the predictions of the Standard Model (SM), such as but not exclusively, apparent lepton flavour universality violation in $B$-meson decays observed in the $b \to c \ell \nu$ transition \cite{Belle:2017rcc,Bobeth:2021lya}, the unsettled anomalous magnetic moment of the muon \cite{Muong-2:2021ojo} and an excess in Dark Matter (DM) searches \cite{Kurinsky:2020dpb,Heikinheimo:2021syx}, all suggesting potential new physics.

Recently, hints of a possible new spin-0 particle, with a mass around 95 GeV, have emerged in multiple independent experiments, such as LEP \cite{ALEPH:2006tnd} in the $e^+e^- \to Z\phi(\to b\overline{b})$ production and decay process (with a 2.3$\sigma$ local excess observed)  and CMS \cite{CMS:2018cyk,CMS:2023yay} when looking for additional Higgs bosons decaying in the $\tau^+\tau^-$ channel (with local significance of 2.6$\sigma$). Additionally, at the same mass value,  ATLAS and CMS reported an anomaly with a local significance of 1.7$\sigma$ and 2.9$\sigma$, respectively,  in the di-photon final state based on their full Run 2 data \cite{ATLAS:2023jzc,CMS:2024yhz}, with a signal strength of  $\mu_{\gamma\gamma}^{\text{ATLAS}}=0.18\pm0.1$ and $\mu_{\gamma\gamma}^{\text{CMS}}=0.33^{+0.19}_{-0.12}$.
These observations, while not yet definitive, have sparked significant theoretical and experimental interest, as they could well point to physics Beyond the SM  (BSM), including extended Higgs sectors or other phenomena. 

In literature, this issue has been addressed in many BSM extensions, such as the 2HDM \cite{Cacciapaglia:2016tlr,Benbrik:2022azi, Benbrik:2022tlg,Azevedo:2023zkg}, the 2HDM plus a real (N2HDM)~\cite{Biekotter:2019kde,Heinemeyer:2021msz,Biekotter:2022jyr} or complex (S2HDM) ~\cite{Biekotter:2023jld,Biekotter:2023oen} singlet, the Next-to-Minimal Supersymmetric SM (NMSSM) \cite{Cao:2016uwt,Domingo:2018uim,Choi:2019yrv,Cao:2019ofo,Biekotter:2021qbc,Cao:2023gkc,Li:2023kbf,Ellwanger:2024vvs,Lian:2024smg,Ellwanger:2024txc,Cao:2024axg}, Left-Right Symmetric Model (LRSM) \cite{Dev:2023kzu}, the Georgi-Machacek (GM) model \cite{Ahriche:2023wkj,Chen:2023bqr,Du:2025eop} and its extensions \cite{Kundu:2024sip}, and other (pseudo)scalar extensions, some of which can explain all these anomalies simultaneously. In fact, there are solutions to the 95 GeV anomalies where a superposition of a CP-even and a CP-odd Higgs boson is also possible to achieve compliance with data, such as in Refs.~\cite{Benbrik:2024ptw,Khanna:2024bah}.

In our study, we investigate to what extent adding an inert doublet to the 2HDM can help address such intriguing anomalies. The resulting model, namely the I(1+2)HDM, is a special case of the 3-Higgs Doublet Models (3HDMs) that have been proposed initially by Ref.~\cite{Keus:2013hya} as a simple extension to 2HDMs with the main purpose of addressing several problems of the latter, including flavour physics, Electro-Weak Symmetry Breaking (EWSB) and DM. 
Adopting  a CP-conserving version of the I(1+2)HDM, which embeds one inert and two active Higgs doublets, we focus on Type-I Yukawa interactions, where all fermions couple to the same Higgs field. This version of the model lends itself well to analytical calculations, featuring a compact parameter space that current and future data can strongly constrain. 
Importantly, the loop effects of the embedded inert charged scalar bosons, $\chi^\pm$, may alter the contributions of the corresponding active states,  $H^\pm$, thereby enhancing or suppressing the di-photon decay rate. A key objective of this paper is to firmly accommodate, within the I(1+2)HDM framework, the most prominent of the aforementioned anomalies, i.e., the one in $\gamma\gamma$ final states--and to explore the parameter space that reproduces it.

The rest of this paper is organised as follows. In Section~\ref{sec:model}, we provide a concise description of the I(1+2)HDM, while Section~\ref{sec:cons} discusses the relevant theoretical and experimental constraints considered in our analysis. Section~\ref{sec:excess} details our numerical methodology for scanning the model's parameter space, aiming to provide a coherent explanation for the three anomalies and to assess their impact on the remaining Higgs spectrum. In Section~\ref{sec:results}, we present our main results and we conclude in Section~\ref{sec:conclusion}.

\section{I(1+2)HDM}
\label{sec:model}
The I(1+2)HDM consists of two active $SU(2)_L$ Higgs doublets with hypercharge $Y=1$ and one inert doublet. Including an additional scalar doublet into the 2HDM has received extensive attention and is considered one of the most thoroughly investigated extensions of the 2HDM~\cite{Grzadkowski:2009bt,Moretti:2015cwa}. The I(1+2)HDM features a discrete $\mathbb{Z}_2 \times \mathbb{Z}^\prime_2$ symmetry, where the first factor corresponds to the inert-doublet $\mathbb{Z}_2$, with only the field $\eta$ transforming as $\eta \rightarrow -\eta$, while $\Phi_{1,2} \rightarrow \Phi_{1,2}$. Furthermore, a softly broken $\mathbb{Z}^\prime_2$ symmetry is introduced for the Higgs doublets, where $\Phi_1$ remains unchanged and $\Phi_2$ is transformed to its negative, $\Phi_2 \rightarrow -\Phi_2$. This symmetry breaking proposed by the Paschos-Glashow-Weinberg theorem \cite{Glashow:1976nt} is necessary to avoid the occurrence of Flavour Changing Neutral Currents (FCNCs) at tree level. 

The most general scalar potential  invariant under both the SM gauge group and the introduced discrete symmetries can be expressed as follows:
\begin{align} 
V(\Phi_1,\Phi_2,\eta) &= -\frac12\left\{m_{11}^2\Phi_1^\dagger\Phi_1 
+ m_{22}^2\Phi_2^\dagger\Phi_2 + \left[m_{12}^2 \Phi_1^\dagger \Phi_2 
+  \text{h.c.} \right] \right\} + m_\eta^2\eta^\dagger \eta \nonumber \\
& + \frac{\rho_1}{2}(\Phi_1^\dagger\Phi_1)^2 
+ \frac{\rho_2}{2}(\Phi_2^\dagger\Phi_2)^2
+ \rho_3(\Phi_1^\dagger\Phi_1)(\Phi_2^\dagger\Phi_2) 
+ \rho_4(\Phi_1^\dagger\Phi_2)(\Phi_2^\dagger\Phi_1) 
\nonumber \\
& + \frac12\left[\rho_5(\Phi_1^\dagger\Phi_2)^2 + \text{h.c.} \right] + \frac{\rho_\eta}{2} 
(\eta^\dagger \eta)^2 + \rho_{1133} (\Phi_1^\dagger\Phi_1)(\eta^\dagger \eta)
+\rho_{2233} (\Phi_2^\dagger\Phi_2)(\eta^\dagger \eta) \nonumber  \\
& +\rho_{1331}(\Phi_1^\dagger\eta)(\eta^\dagger\Phi_1) 
+\rho_{2332}(\Phi_2^\dagger\eta)(\eta^\dagger\Phi_2) 
+\frac{1}{2}\left[\rho_{1313}(\Phi_1^\dagger\eta)^2 +\text{h.c.} \right] \nonumber  \\
&  +\frac{1}{2}\left[\rho_{2323}(\Phi_2^\dagger\eta)^2 +\text{h.c.} \right].
\label{pot} 
\end{align}
Here, $\rho_i$ represent the quartic coupling parameters, while $m_\eta^2$, $m_{11}^2$, $m_{22}^2$ and $m_{12}^2$ are mass-squared terms. The values of $m_{22}^2$ and  $m_{12}^2$ are determined by the conditions for minimising the potential, with all parameters required to be real in order to ensure CP-conservation. 

For a region of parameter space where EWSB may occur, the minimisation of this potential gives
\begin{equation} 
\Phi_1=\left(
\begin{array}{c}
0 \\ 
v_1/\sqrt{2}
\end{array}\right), \quad
\Phi_2=\left(
\begin{array}{c}
0 \\ 
v_2/\sqrt{2}
\end{array}
\right),
\label{eq:ewsb}
\end{equation}
where $v_1$ and $v_2$ are Vacuum Expectation Values (VEVs) of the neutral component of each (pseudo)scalar doublet $\Phi_i$, while the $\eta$ scalar doublet remains inert due to the $\mathbb{Z}_2$ symmetry, which ensures that its constituent (pseudo)scalar states do not mix with those of $\Phi_1$ and $\Phi_2$. 
\\
So that, with three complex scalar $SU(2)_L$ doublets, there are twelve fields:
\begin{equation} 
\Phi_1=\left(
\begin{array}{c}
\phi_1^\pm   \\ 
(v_1+\eta_1+i z_1)/\sqrt{2}
\end{array}\right), 
\quad
\Phi_2=\left(
\begin{array}{c}
\phi_2^\pm \\ 
(v_2+\eta_2+i z_2)/\sqrt{2}
\end{array}
\right),  \quad
\eta = \left(
\begin{array}{c}
 \chi^\pm \\ 
 (\chi + i \chi_a)/\sqrt{2} 
\end{array}
\right).
\label{doublets}
\end{equation}
Therefore, apart from the three fields that are used to provide mass to the weak gauge bosons ($W^\pm$ and $Z$), the remaining nine fields are physical (pseudo)scalar states. Here, we distinguish between two independent spectra.
\begin{itemize}
\item [$\circ$] The first one reflects the pure 2HDM Higgs spectrum that consists of two CP-even Higgs bosons, $h$ and $H$, one CP-odd state $A$ and two charged Higgs bosons $H^{\pm}$. The corresponding mass terms of these particles are denoted as $m_h$, $m_H$ (with $m_h < m_H$), $m_A$ and $m_{H^{\pm}}$, respectively, and can be expressed in terms of the parameters of the potential and {VEV}s. The mass-squared matrix of the CP-even scalar sector can be diagonalised through the angle $\alpha$  whereas both the CP-odd and charged sectors share the same mixing angle, defined by the ratio of the two VEVs of the active Higgs fields, $\tan\beta= v_2/v_1$. For more details, we refer the reader to Refs.~\cite{Branco:2011iw,Gunion:2002zf}.
\item [$\circ$] In addition, the inert sector is made up by three (pseudo)scalar particles, $\chi$, $\chi_a$ and $\chi^{+}$, with squared masses given by (here, $v^2=v_1^2+v_2^2=(246~{\rm GeV})^2$)
\begin{eqnarray}
&& m^2_{\chi^{\pm}} = m_{\eta}^2 + \frac{1}{2} \rho_{a} v^2, \nonumber\\
&& m^2_{\chi} = m^2_{\chi^{\pm}} + \frac{1}{2} (\rho_{b} + \rho_{c})v^2,\nonumber\\
&& m^2_{\chi_a} = m^2_{\chi^{\pm}} + \frac{1}{2} (\rho_{b} - \rho_{c})v^2,
\end{eqnarray}
wherein the so-called `dark democracy' approach \cite{Keus:2015hva,Cordero-Cid:2016krd,Cordero-Cid:2018man,Dey:2024epo} has been adopted primarily to reduce the number of free parameters in the model, thereby making the analysis more tractable while still preserving the essential features of the I(1+2)HDM. This approach allows us to redefine the new couplings as follows:
\begin{align} \label{Eq:DarkDemocracy}
\rho_a\equiv \rho_{1133}&=\rho_{2233}, \nonumber \\
\rho_b\equiv \rho_{1331}&=\rho_{2332}, \nonumber \\ 
\rho_c\equiv \rho_{1313}&=\rho_{2323} .
\end{align} 
and hence the condition
\begin{equation}
\label{eq:massChi}
m^2_{\chi}=m_\eta^2 + \frac{v^2}{2}(\rho_a+\rho_b+\rho_c) > 0,
\end{equation}
must be met, ensuring that the inert doublet does not develop a tachyonic mass.
\end{itemize}

Thus, we are left with 12 free parameters,
\begin{equation}
\Sigma = \left\{m_h,\, m_A,\, m_H,\, m_{H^\pm},\, m_{12}^2,\, \tan\beta,\, \sin\alpha,\, m_{\chi},\, m_{\chi_a},\, m_{\chi^\pm},\, m_{\eta}^2,\, \rho_{\eta} \right\}, 
\label{set}
\end{equation}
which we use then 
to describe the Higgs sector of the I(1+2)HDM, while considering $H$ as SM-like (i.e., $m_H\approx 125$ GeV).

\noindent
Furthermore, following the enforcement of the symmetries described above, the most general Yukawa interactions in the I(1+2)HDM  remain similar to those of the 2HDM and the corresponding Lagrangian reads as    
\begin{align}
\mathcal{L}_{\rm Yuk}^{{\rm I(1+2)HDM}} = -{\overline Q}_L Y_u \widetilde{\Phi}_uu_R^{} - {\overline Q}_L Y_d \Phi_dd_R^{} - {\overline L}_LY_\ell \Phi_\ell \ell_R^{} + {\rm h.c.} \supset  - \sum_{f=u,d,\ell} \frac{m_f}{v} \big( \xi_h^f \bar{f}f h + \xi_H^f \bar{f}f H \big),
\end{align}
where $Y_{f}$ ($f=u$, $d$ or $l$) are  $3\times 3$ Yukawa matrices and $\tilde{\Phi}_{1,2}=i\sigma_{2}\Phi_{1,2}^*$, with $\sigma_{1,2}$ being the Pauli matrices.
In Tab.~\ref{table1} we list all the Type-I Yukawa couplings $\xi_h^f $ and $\xi_H^f $ of the CP-even Higgs bosons, $h$ and $H$.

\begin{table}[ht]
\vspace{0.15cm}
\centering
\begin{tabular}{cccccc}
\hline\hline                        
$\xi_h^u$  &  $\xi_h^d$    &  $\xi_h^\ell$   &  $\xi_H^u$   &   $\xi_H^d$  &   $\xi_H^\ell$  \\ [0.5ex]  
\hline 
$\cos\alpha/\sin\beta$  &  $\cos\alpha/\sin\beta$    &   $\cos\alpha/\sin\beta$   &  $\sin\alpha/\sin\beta$   &   $\sin\alpha/\sin\beta$  &   $\sin\alpha/\sin\beta$ \\
\hline
\end{tabular}
\caption{Type-I Yukawa couplings  of the neutral Higgs bosons $h,\,H$ 
to the up-quarks, down-quarks and leptons in the I(1+2)HDM, normalised by the corresponding SM Yukawa couplings.}
\label{table1}   
\end{table}

\section{Theoretical and Experimental Constraints}
\label{sec:cons}
For our randomly generated scan, we considered only points that are physically viable in the sense that they obey theoretical and experimental requirements. Here, we present a summary of the involved constraints imposed on the I(1+2)HDM Type-I. 
\subsection{Theoretical Constraints} 
These are the following ones.
\begin{itemize}
\item [$\circ$]  \textbf{Perturbativity}: We require that all quartic dimensionless couplings $\rho_i$ in Eq.~(\ref{pot}) remain ${\le 4\pi}$ to avoid a non-perturbative behaviour. 
\item [$\circ$]  \textbf{Unitarity}: To satisfy unitarity, it is necessary for the magnitudes of the eigenvalues of matrices, which are formed in the bases of different $2 \rightarrow 2$ scalar scattering states, to be limited to a maximum value of $8\pi$. (A comprehensive examination of unitarity in 3HDMs can be found in~\cite{Moretti:2015cwa}.)
\item [$\circ$] \textbf{Vacuum Stability}: To maintain the stability of the Higgs potential, it is necessary to ensure that the latter remains bounded from below in all directions in field space. In Ref.~\cite{Grzadkowski:2009bt}, the necessary and sufficient conditions to ensure the potential remains positive have been established:
\begin{align} 
\rho_1&>0,\quad \rho_2>0,\quad \rho_\eta>0,\\
\rho_{x} &>-\sqrt{\rho_1\rho_2},
\quad \rho_{y}>-\sqrt{\rho_1\rho_\eta},\quad 
\rho_{y}>-\sqrt{\rho_2\rho_\eta},\\
\rho_{y} &\geq0 \vee \left(\rho_\eta \rho_{x}-\rho_{y}^2
>-\sqrt{(\rho_\eta\rho_1-\rho_{y}^2)
(\rho_\eta\rho_2-\rho_{y}^2)}\right),
\label{constraints}
\end{align}
where 
\begin{eqnarray}
\rho_{x}&=&\rho_3+\min\left(0,\rho_4-|\rho_5|\right),\\
\rho_{y}&=&\rho_{a}+\min\left(0,\rho_{b}-|\rho_{c}|\right).
\end{eqnarray}
\end{itemize}

\subsection{Experimental Constraints} 
Experimental constraints provide the most accurate information about the Higgs mass spectrum, whether for the observed one at 125 GeV or companion states possibly existing at the EW scale. In this regard,  EW Precision Observables (EWPOs), SM-like Higgs boson measurements, new Higgs boson direct searches as well as  flavour observables set limits on the various Higgs boson properties. Below, we summarise the most important constraints.
\begin{itemize}
\item [$\circ$] \textbf{EWPOs}:
The precision tests of quantum effects on EW parameters, such as gauge couplings and gauge boson masses, are quite important to check the validity of any new BSM physics. Thus, $S$, $T$ and $U$, the so-called `oblique parameters'~\cite{Peskin:1991sw,Grimus:2008nb} that characterise EW radiative corrections, need to be compared to the ensuing I(1+2)HDM predictions. By fixing $U=0$\footnote{This choice is motivated by the fact that $U$ is suppressed by an additional factor $m_Z^2/m_H^2$ compared to both $S$ and $T$, also it's considered a simplifying approach in scalar-sector extensions. In our scenario, the contributions to $U$ remains typically subdominant compared to others oblique parameters, due to the relatively small allowed mass splitting between the inert and active doublets.}, the currently measured values~\cite{ParticleDataGroup:2024cfk} of the remaining observables are:
\begin{eqnarray}
 \Delta S = -0.05 \pm 0.07, \quad
 \Delta T = 0.00 \pm 0.06 \quad {\rm and} \quad \rho_{ST} = 0.93~({\rm with}~\Delta U=0), 
\end{eqnarray} 

where
\begin{equation}
\Delta X^{\text{I(1+2)HDM}} = \Delta X^{\text{2HDM}} + \Delta X^{\text{IDM}}, \quad X= S, T. 
\end{equation}
The formulae for the 2HDM  can be found in Refs.~\cite{Grimus:2008nb,Grimus:2007if,Peskin:1991sw} and \cite{Barbieri:2006dq}, respectively. In Ref.~\cite{Moretti:2015cwa}, the authors presented the general expressions for the I(1+2)HDM.
\item [$\circ$] \textbf{SM-like Higgs Boson Discovery}: 
To ensure the consistency between our parameter space and the experimental measurements of the properties of the discovered SM-like Higgs boson with a mass of approximately 125 GeV, we have utilised the publicly available code \texttt{HiggsSignals-3} \cite{Bechtle:2020uwn}. This code allows us to evaluate a  $\chi^2$ measure, which assesses the compatibility of the Higgs signal strengths observed at the Tevatron and LHC with the predictions of our model. We have selected the I(1+2)HDM parameter space that satisfies the condition $\chi^2_{125}-{\rm min}\big(\chi^2_{125}\big) \le {\rm 6.18}$, corresponding to a Confidence Level (C.L.) of 95\% with 159 degrees of freedom, where $\chi_{125}^2$  represents the value computed by \texttt{HiggsSignals-3} for the 125 GeV Higgs signal strength measurements. More importantly, in order to assess the performance of the I(1+2)HDM relative to the SM, we have explicitly computed the difference $\Delta\chi^2_{125} = \chi^2_{\rm SM} - {\rm min}\big(\chi^2_{\rm I(1+2)HDM}\big) = 152.54-159.36$=-6.82. This indicates that the I(1+2)HDM fit is slightly worse than the SM in term of the 125 GeV Higgs signal strengths. However, the resulting value $\chi^2_{\rm min}/{\it ndf}=159.36/159=1.002$ still corresponds to an excellent global fit, consistent with current Higgs precision 
data\footnote{Here, {\it ndf} stands for the number of degrees of freedom.}.
\item [$\circ$]  \textbf{Non-SM-like Higgs Boson Exclusions}: 
To ensure that our parameter space remains within the acceptable exclusion limits from new Higgs boson searches carried out at LEP, the Tevatron and LHC, {we have used  the \texttt{HiggsTools} package \cite{Bahl:2022igd}, which embeds the most recent \texttt{HiggsBounds-6} \cite{Bechtle:2020pkv,CMS:2024ulc} dataset, such that only parameter points with an observed limit ratio below 0.95 for each scalar (i.e., below the 95$\%$ C.L. experimental exclusion) are retained.}
\item [$\circ$]  \textbf{Flavour Constraints}: To test the I(1+2)HDM model against flavour physics constraints, we have used the public {\tt C++} {\tt SuperIso} \cite{Mahmoudi:2008tp} program, and so several observables have been checked such as: $BR(B \to X_s \gamma)$ \cite{HFLAV:2016hnz}, BR$(B_s \to \mu^+\mu^-)$ \cite{LHCb:2021awg,LHCb:2021vsc}, BR$(B \to \tau\nu)$\cite{HFLAV:2016hnz} and many others.
\end{itemize}	

\section{The Excesses in $h \to \gamma\gamma,~\tau^+\tau^-$ and $b\bar{b}$ Channels}
\label{sec:excess} 
Our analysis involves evaluating the signal strengths of the aforementioned excesses within the I(1+2)HDM (we will use the notation NP in formulae to refer to it) using the Narrow Width Approximation (NWA) and examining whether a unified explanation for all these may be achieved. The results of searches for a possible 95 GeV new state, that we denote by $\phi_{95}^{\rm NP}$ and assume to be CP-even (i.e., a scalar), as mentioned, can be cast in terms of signal strengths relative to the predictions for a SM-like Higgs state of 95 GeV (hereafter, denoted by $h_{95}^{\rm SM}$) in each of the discussed three channels. 

\begin{itemize}
\item [$\circ$] \textbf{Searches in the $b\bar{b}$ channel}: 
\begin{equation}
\mu_{b\bar{b}}(\phi_{95})=\frac{\sigma(e^+e^- \to Z \phi_{95}^{\rm NP}) \text{BR}(\phi_{95}^{\rm NP} \to b\bar{b})}{\sigma(e^+e^- \to Z h_{95}^{\rm SM}) \text{BR}(h_{95}^{\rm SM} \to b\bar{b})}=\left|c_{\phi_{95}^{\rm NP}ZZ}\right|^2 \times \frac{\text{BR}(\phi_{95}^{\rm NP} \to b\bar{b})}{\text{BR}(h_{95}^{\rm SM} \to b\bar{b})}.
\label{eq:mubb}
\end{equation}
\item [$\circ$] \textbf{Searches in the $\gamma\gamma$ channel}: 
\begin{equation}
\mu_{\gamma\gamma}(\phi_{95})=\frac{\sigma(gg \to \phi_{95}^{\rm NP}) \text{BR}(\phi_{95}^{\rm NP} \to \gamma\gamma)}{\sigma(gg \to h_{95}^{\rm SM}) \text{BR}(h_{95}^{\rm SM} \to \gamma\gamma)}=\left|c_{\phi_{95}^{\rm NP}t\bar{t}}\right|^2 \times \frac{\text{BR}(\phi_{95}^{\rm NP} \to \gamma\gamma)}{\text{BR}(h_{95}^{\rm SM} \to \gamma\gamma)}.
\label{eq:mugaga} 
\end{equation}
\item [$\circ$] \textbf{Searches in the $\tau^+\tau^-$ channel}: 
\begin{equation}
\mu_{\tau^+\tau^-}(\phi_{95})=\frac{\sigma(gg \to \phi_{95}^{\rm NP}) \text{BR}(\phi_{95}^{\rm NP} \to \tau^+\tau^-)}{\sigma(gg \to h_{95}^{\rm SM}) \text{BR}(h_{95}^{\rm SM} \to \tau^+\tau^-)}=\left|c_{\phi_{95}^{\rm NP}t\bar{t}}\right|^2 \times \frac{\text{BR}(\phi_{95}^{\rm NP} \to \tau^+\tau^-)}{\text{BR}(h_{95}^{\rm SM} \to \tau^+\tau^-)}.
\label{eq:mutautau}
\end{equation}
\end{itemize}
Here, $c_{\phi_{95}^{\rm NP}ZZ}$ and $c_{\phi_{95}^{\rm NP}t\bar{t}}$ are the reduced couplings of the $\phi_{95}^{\rm NP}$ to $ZZ$ and $t\bar t$ compared to the SM values, respectively. 

With a local significance of $2.3\sigma$ and $3.1\sigma$, respectively, at a mass value around 95 GeV, the four LEP Collaborations \cite{LEPWorkingGroupforHiggsbosonsearches:2003ing} and CMS\cite{Biekotter:2023oen} have reported the following experimental values for the $b\bar{b}$ and $\tau^+\tau^-$ signal strengths:
\begin{equation}
\label{eq:exp_values_bb_tautau}
\mu_{b\bar{b}}^{\rm LEP}(\phi_{95}) = 0.117 \pm 0.057 \quad \text{and} \quad \mu_{\tau^+\tau^-}^{\rm CMS}(\phi_{95}) = 1.2 \pm 0.5,
\end{equation}
while for the di-photon channel, a local excess of $1.7\sigma$ and $2.9\sigma$ is reported by both ATLAS\cite{Biekotter:2023oen},  $\mu_{\gamma\gamma}^{\rm ATLAS}(\phi_{95}) = 0.18 \pm 0.1$, and CMS\cite{CMS:2024yhz},  $\mu_{\gamma\gamma}^{\rm CMS}(\phi_{95}) = 0.33^{+0.19}_{-0.12}$ (with the peak position between 95.3 and 95.4 GeV).
Recently, a combined signal strength of the two experimental results in this channel, without possible correlations, was given as \cite{Biekotter:2023oen}
\begin{equation}
\label{eq:exp_value_gamgam}
\mu_{\gamma\gamma}^{\rm ATLAS+CMS}(\phi_{95}) = 0.24^{+0.09}_{-0.08}.
\end{equation}
Consequently, in the following, we examine whether these anomalies can be accommodated within the  parameter space  of the I(1+2)HDM, focusing our discussion on the Type-I scenario. To do so, a $\chi^2$ analysis is performed for each channel 
involving the $h\equiv \phi_{95}^{\rm NP}$ state as follows:
\begin{eqnarray}
\chi^2_{\gamma\gamma,\tau^+\tau^-,b\bar{b}}=\frac{\left(\mu_{\gamma\gamma,\tau^+\tau^-,b\bar{b}}-\mu_{\gamma\gamma,\tau^+\tau^-,b\bar{b}}^\mathrm{ exp}\right)^2}{\left(\Delta\mu^\mathrm{exp}_{\gamma\gamma,\tau^+\tau^-,b\bar{b}}\right)^2}, 
\end{eqnarray}
where we consider only points that lie within the 95$\%$ C.L. region. However, to make sure that such a sample not only explains the observed excesses but also remains consistent with the well-measured proprieties of $h_{125}$, we also define
\begin{eqnarray}
\chi^2=\chi^2_{125}  + \chi^2_{\gamma\gamma,\tau^+\tau^-,b\bar{b}}.
\end{eqnarray}

\section{Results}
\label{sec:results}
To perform our numerical analysis, we have implemented the I(1+2)HDM Type-I in a {\tt Fortran} code which allows a fast numerical evaluation of the relevant experimental observables over the theoretical parameter space. The data generated are then tested against the constraints previously discussed, {through a {\tt Python} interface to the \texttt{HiggsTools} package}. As mentioned, in doing so, the heaviest CP-even Higgs boson, $H$, is set to mimic the SM-like Higgs boson observed at the LHC,
specifically,  with a mass of  125.09 GeV. Crucially, the mass of the lightest CP-even Higgs, $h$, is allowed to vary within the range $[94,\,97]$ GeV, where the $\gamma\gamma$ excess is most pronounced  (as emphasised in the introduction). The remaining scanned parameters are summarised in Tab.~\ref{tab:par-scan}.  

\begin{table}[!h]
\centering
\begin{tabular}{lcccccccccc}
\toprule[1pt]
Parameter & $m_h$ & $m_H$ & $m_A$ & $m_{H^\pm}$ & $\tan{\beta}$ \\
\midrule[1pt]
Scan Range &  $[94,\,97]$ & $125.09$  &  $[100,\,10^3]$   &  $[90,\,10^3]$  & $[0.5,\,50]$ \\
\bottomrule[1pt]
\toprule[1pt]
Parameter &  $\sin({\beta - \alpha})$ &  $m_{12}^2$ & $m_{\chi},m_{\chi_a},m_{\chi^{\pm}}$ & $\rho_{\eta}$ & $m_{\eta}^2$\\
\midrule[1pt]
Scan Range &  $[-0.4 ,\, 0.5]$  &  $[-10^3 ,\, 10^3]$  & $[90 ,\, 10^3]$  &  $[7 ,\, 12]$  &  $[-10^5 ,\, 10^5]$ \\
\bottomrule[1pt]
\end{tabular}
\caption{Scan ranges of the I(1+2)HDM Type-I  parameters. (Mass (squared) are in GeV$^{(2)}$.)}
\label{tab:par-scan}
\end{table}

All excesses are then evaluated simultaneously using a $\chi^2$ test, requiring $\chi^2_{\gamma\gamma+\tau^+\tau^-+b\bar{b}} \leq 8.03$ at the $2\sigma$ C.L. to ensure compatibility with data. We anticipate here that our results will point towards the possible existence of a light CP-even (or scalar) Higgs boson, supporting our exploration of the I(1+2)HDM Type-I. These will be mapped in the forthcoming plots in terms of two-dimensional planes of the signal strength parameters: ($\mu_{bb}$ - $\mu_{\gamma\gamma}$), ($\mu_{\tau^+\tau^-}$ - $\mu_{\gamma\gamma}$) and ($\mu_{\tau^+\tau^-}$ - $\mu_{bb}$), thereby providing insights into parameter correlations and the overall model fit.   
\begin{figure}[!t]
\centering
\includegraphics[width=0.325\textwidth]{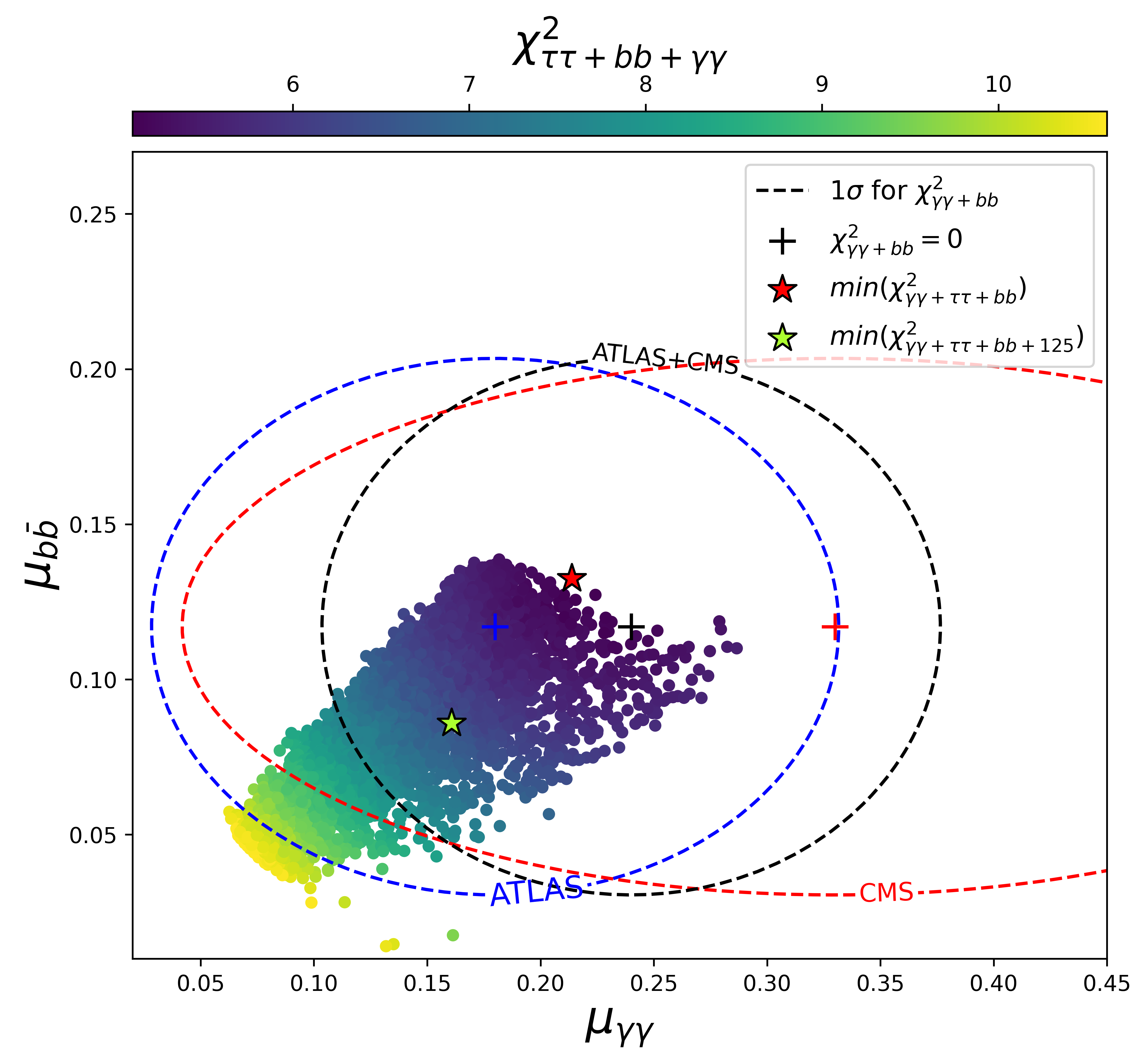}
\includegraphics[width=0.325\textwidth]{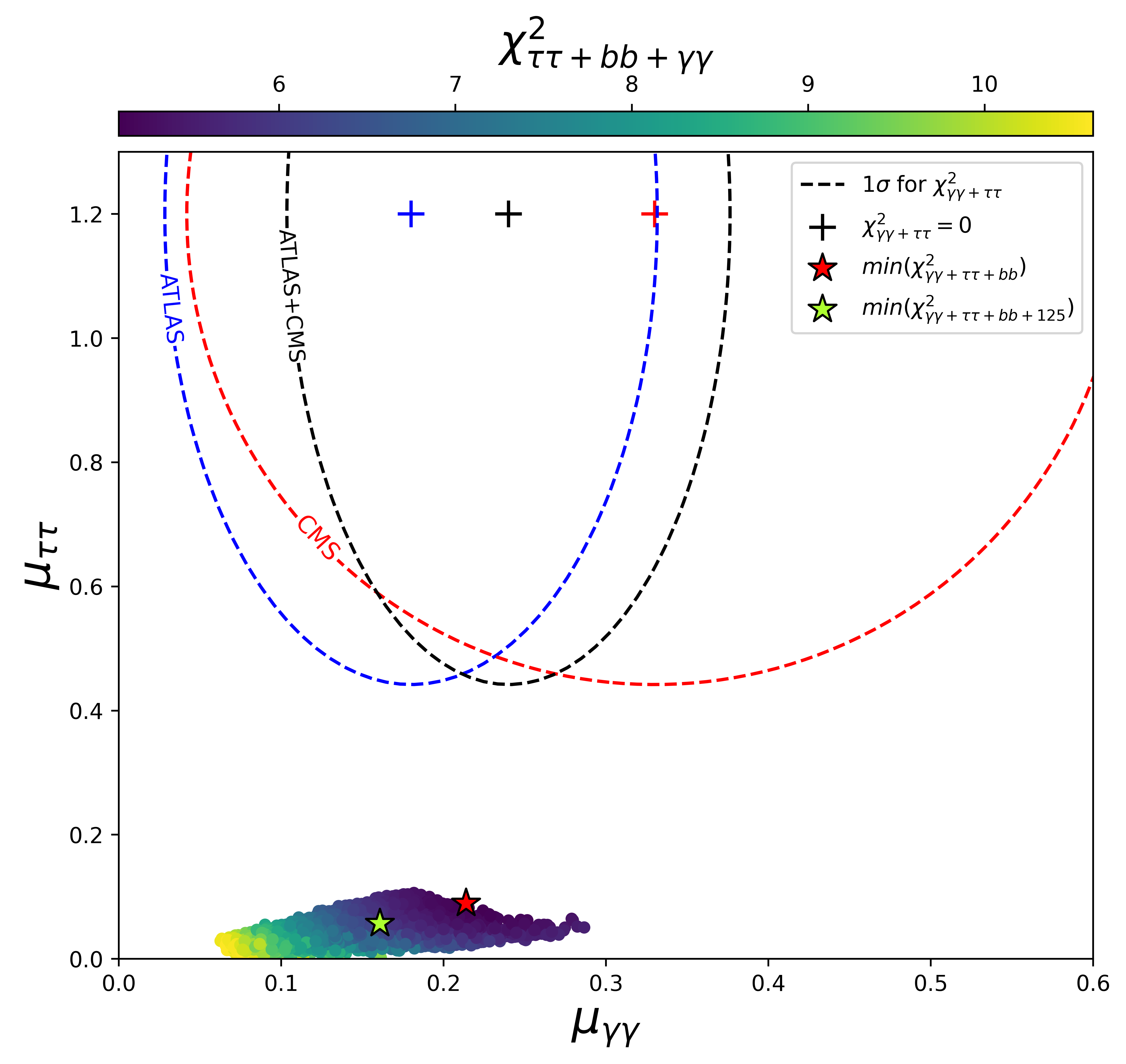}
\includegraphics[width=0.325\textwidth]{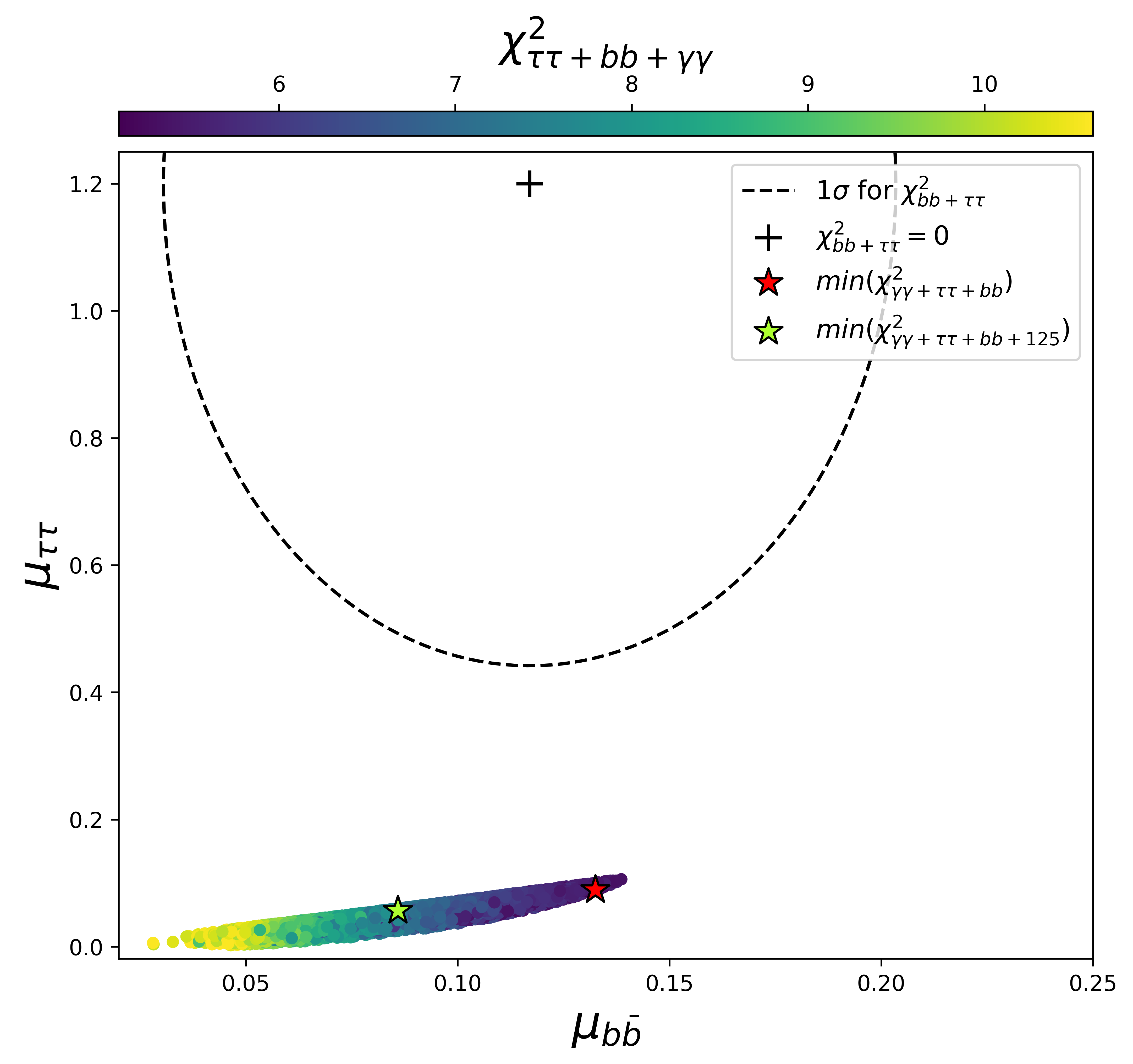}\\
\includegraphics[width=0.328\textwidth]{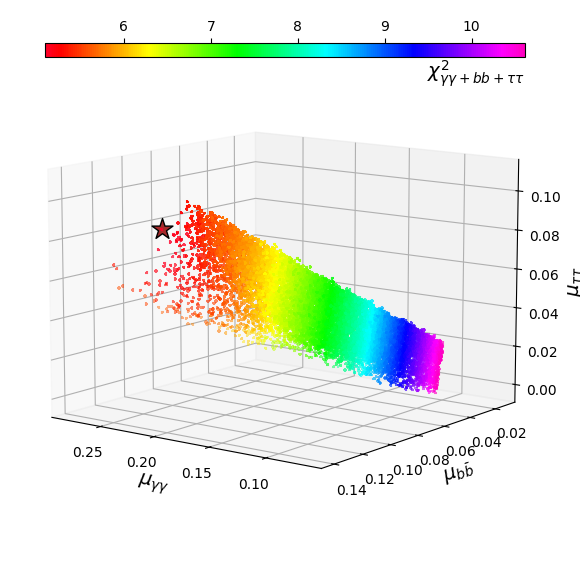}
\includegraphics[width=0.328\textwidth]{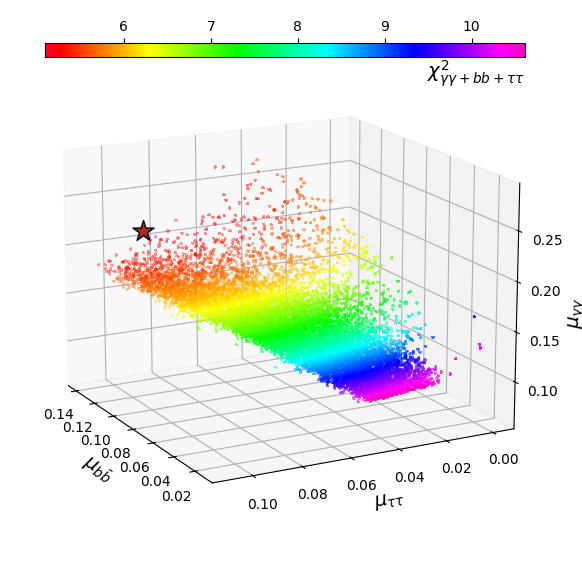}
\includegraphics[width=0.328\textwidth]{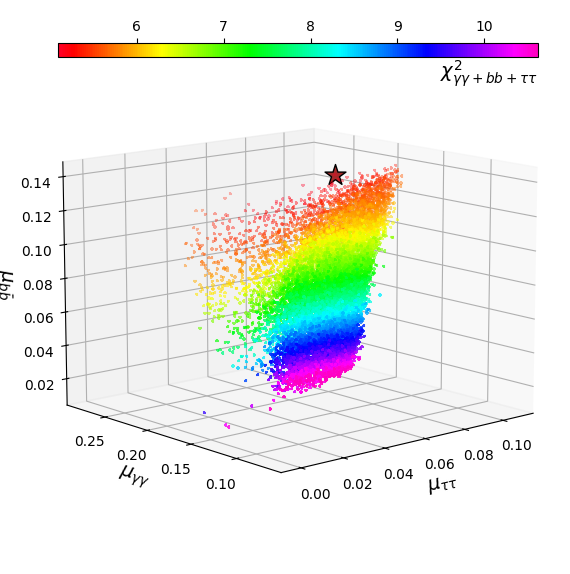}
\caption{Correlations among the signal strengths in Eqs.~(\ref{eq:mubb})--(\ref{eq:mutautau}), with the $\chi^2_{\gamma\gamma+\tau^+\tau^-+b\bar{b}}$ value represented by the colour bar. The red star marks the $\min(\chi^2_{\gamma\gamma+\tau^+\tau^-+b\bar{b}})$ best-fit point, while the green star indicates the $\min(\chi^2_{\gamma\gamma+\tau^+\tau^-+b\bar{b}+125})$ best-fit point including the SM-like Higgs data. The ATLAS and CMS results, along with their corresponding 1$\sigma$ bands, are also depicted through the ellipse contours, which indicate the regions corresponding to the excess observed at such a  C.L. (We use the $\chi^2_x+\chi^2_y=2.30$ relation, where the indexes $x$ and $y$ represent each possible pair of the three signal channels: $\gamma\gamma$, $\tau^+\tau^-$ and $b\bar{b}$.)}
\label{fig1}
\end{figure}

Fig.~\ref{fig1} displays the results for our fits 
as a colour map, projected onto the signal strength planes ($\mu_{\gamma\gamma}$ - $\mu_{bb}$ )(left), ($\mu_{\gamma\gamma}$ - $\mu_{\tau^+\tau^-}$) (middle) and ($\mu_{bb}$ - $\mu_{\tau^+\tau^-}$) (right), following our scan of I(1+2)HDM Type-I parameter space. The black cross corresponds to $\chi^2_{x,y}=0$ (for all combinations of $x,y=\gamma\gamma, \tau^+\tau^-, b\bar b$) whereas the green and red stars mark the points of ${\rm min}(\chi^2_{\gamma\gamma+\tau^+\tau^-+b\bar{b}+125})$ and ${\rm min}(\chi^2_{\gamma\gamma+\tau^+\tau^-+b\bar{b}})$, representing the best-fit point with and without including the SM-like Higgs data in the fit, respectively. The clear clustering shown in the upper left panel predicts a correlation between $\mu_{\gamma\gamma}$ and $\mu_{bb}$, with some spread. Most allowed points cluster around $\mu_{bb}\sim0.1$ and $\mu_{\gamma\gamma}\sim0.1-0.25$, appearing overall quite consistent with the experimental data, as reflected by the location of the best-fit points. In contrast, a moderate values for $\mu_{\tau^+\tau^-}$ ($<0.2$) are more challenging but still achievable in part of the parameter space as can be seen from the upper middle panel, reflecting the stringent experimental constraints in the $\tau^+\tau^-$ channel. It is straightforward to see, from both plots, that the combined $1\sigma$ contour encompasses the $\chi^2$ minima, consistent with the interpretation of a  potential light Higgs boson. The situation does not change substantially in the upper right panel, where a positive correlation between $\mu_{b\bar b}$ and $\mu_{\tau^+\tau^-}$ is highlighted. So, fitting simultaneously large $\mu_{b\bar b}$ and $\mu_{\tau^+\tau^-}$ rates might not be trivial.

Still in Fig.~\ref{fig1}, to guide further investigations, we present our results for $\chi^2_{\gamma\gamma+\tau^+\tau^-+b\bar{b}}$ by displaying its colour map projected onto  a three-dimensional (3D) space, where the axes represent  $\mu_{b\bar b}$, $\mu_{\tau^+\tau^-}$ and $\mu_{\gamma\gamma}$. It becomes clear that all the three visualisations illustrate the strong correlation among the signal strengths required to fit the excesses, highlighting non-trivial yet viable regions in parameter space where the I(1+2)HDM Type-I provides a good fit to explain the experimental anomalies.

Building on the above findings, we notice that the inert charged states  $\chi^\pm$ play a crucial role in determining $\mu_{\gamma\gamma}$ and, consequently, influence all other observables. Specifically, the decay $h \to \gamma\gamma$ proceeds at one-loop level through the virtual exchange of $\chi^\pm$, in addition to the 2HDM-like contribution from $H^\pm$ and the SM contribution dominated by the $W^\pm$ and top loops. The corresponding decay width is explicitly given by 
\begin{align}
\Gamma(h \to \gamma\gamma)=& \frac{G_F\,\alpha^2\,m_h^3}{128\sqrt{2}\pi^3}\bigg|\sum_{f}\underbrace{Q_f^2 N_c\,\eta_{hff}\,A^{\gamma \gamma}_{\frac{1}{2}}(\tau_f)}_{C_{\rm top}}+ \underbrace{\eta_{hWW}\,A^{\gamma \gamma}_{1}(\tau_W)}_{C_W}  \underbrace{-\frac{m_W}{g\,m_{H^\pm}^2}\,\eta_{hH^{\pm}H^{\mp}}\,A_{0}^{\gamma \gamma}(\tau_{H^\pm})}_{C_{H^\pm}} \nonumber                                                   \\
 & \underbrace{-\frac{m_W}{g\,m_{\chi^\pm}^2}\,\eta_{h\chi^{\pm}\chi^{\mp}}\,A_{0}^{\gamma \gamma}(\tau_{\chi^\pm})}_{C_{\chi^\pm}} \bigg|^2.
\label{width:h2gaga}                                                                                                                                                                                                                                                                                                                                      
\end{align}
Here, $N_c$ represents the colour factor, with $N_c=3$ for quarks and $N_c=1$ for leptons, while $Q_f$ denotes the electric charge of a particle  in the loop. The fine-structure constant is denoted by $\alpha$, while $A^{\gamma\gamma}_{1/2}$, $A^{VV}_{1}$ and $A^{\gamma\gamma}_{0}$ are the form factors associated with: spin-$1/2$ fermions ($f$), $W^\pm$ boson and charged scalars ($H^\pm$ and $\chi^\pm$), respectively. Such dimensionless functions characterise the contributions of each particle and can be expressed in terms of  Passarino-Veltman functions \cite{Passarino:1978jh}.

\begin{figure}[!hb]
\centering
\hspace{-0.02\textwidth}
\includegraphics[width=0.328\textwidth]{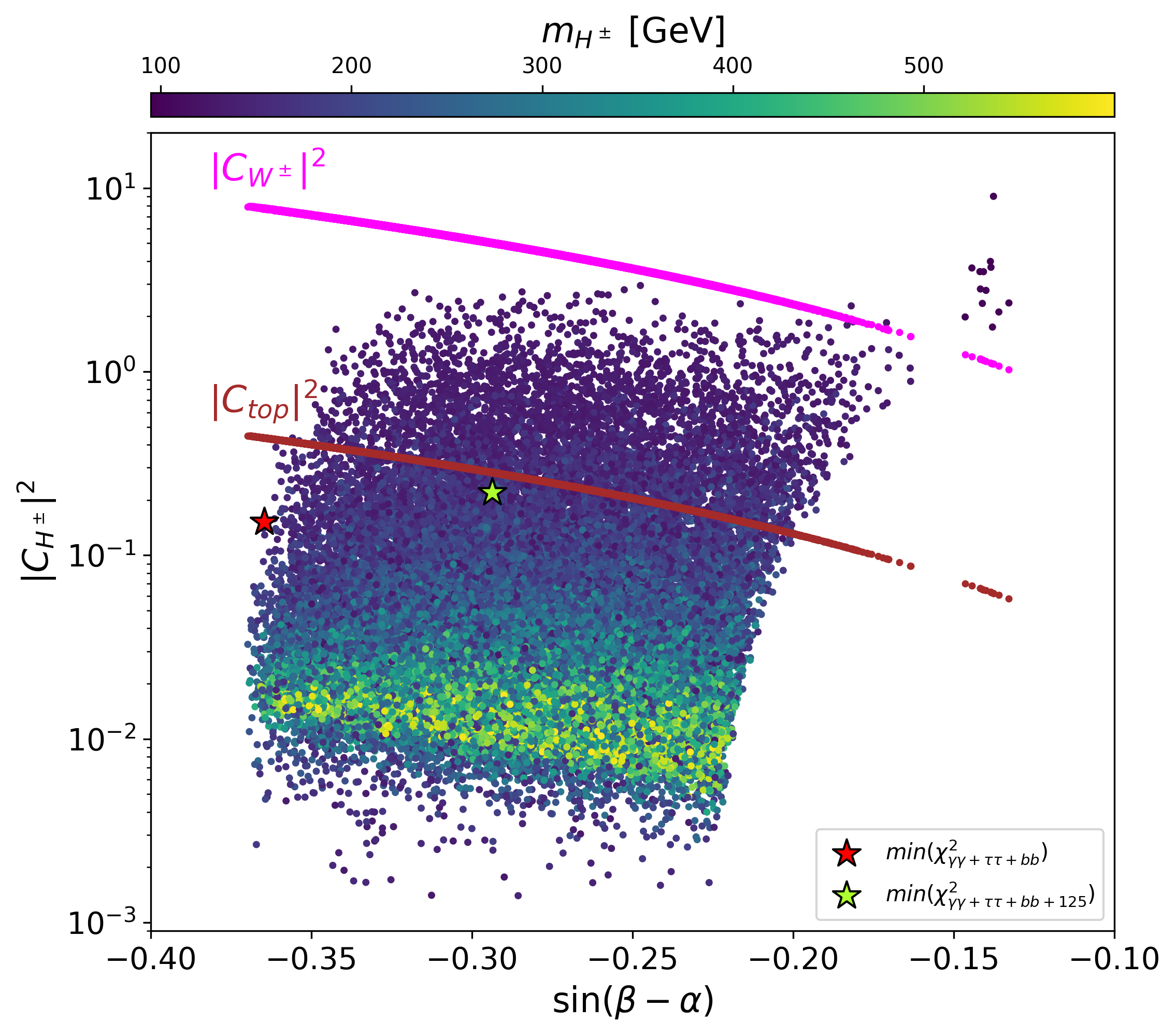}
\includegraphics[width=0.328\textwidth]{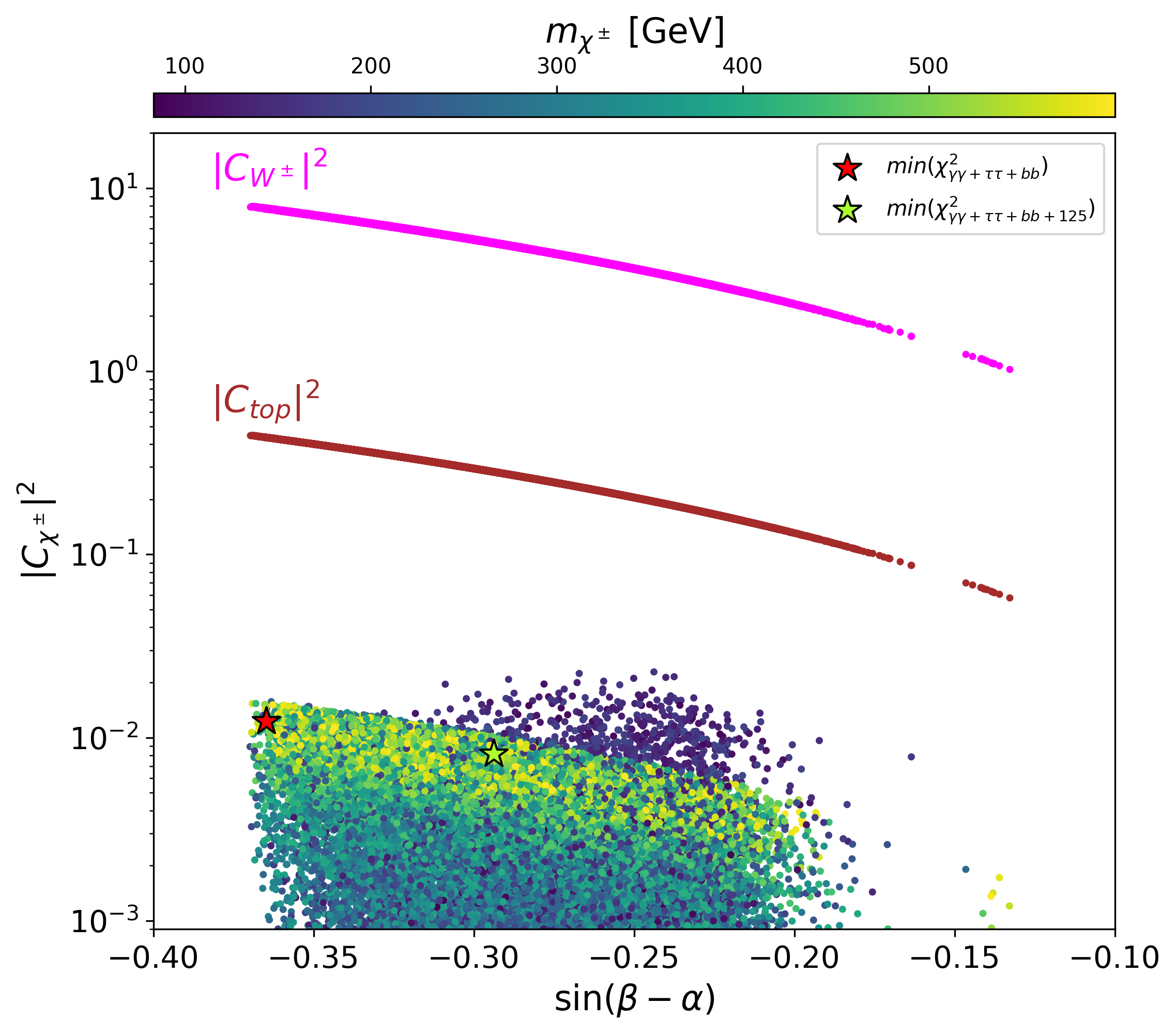}\\[1em]
\includegraphics[width=0.328\textwidth]{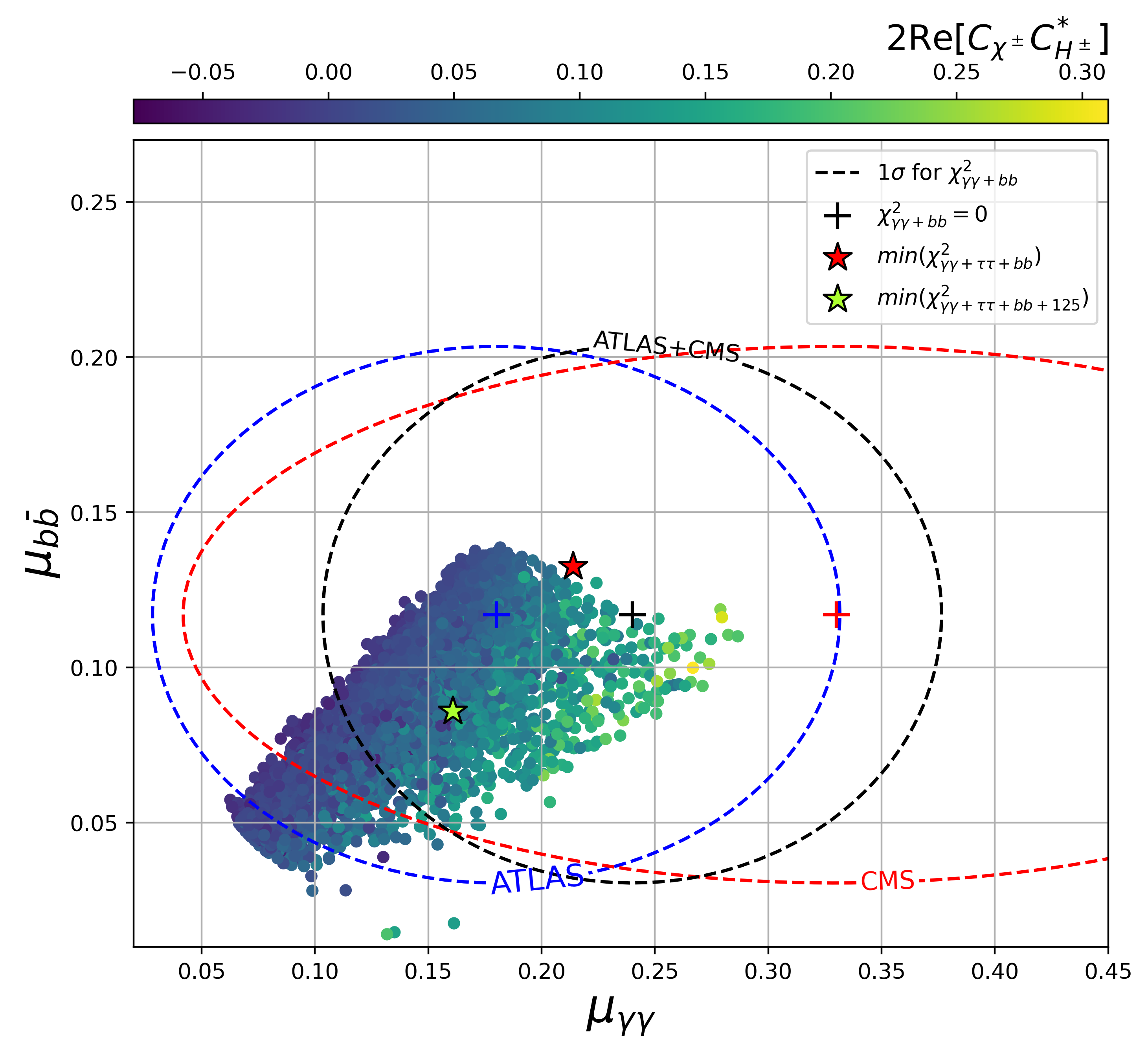}
\includegraphics[width=0.328\textwidth]{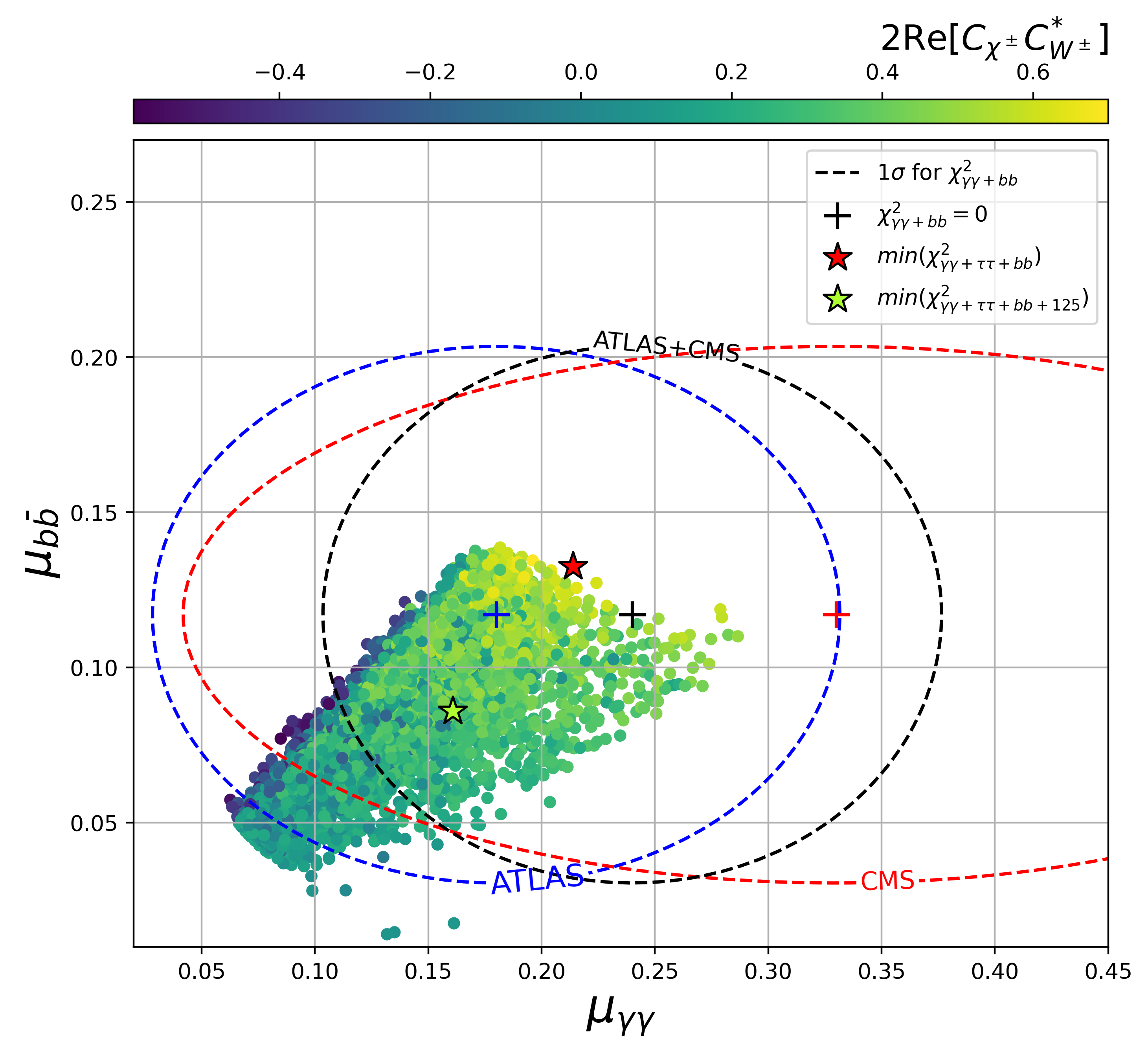}
\includegraphics[width=0.328\textwidth]{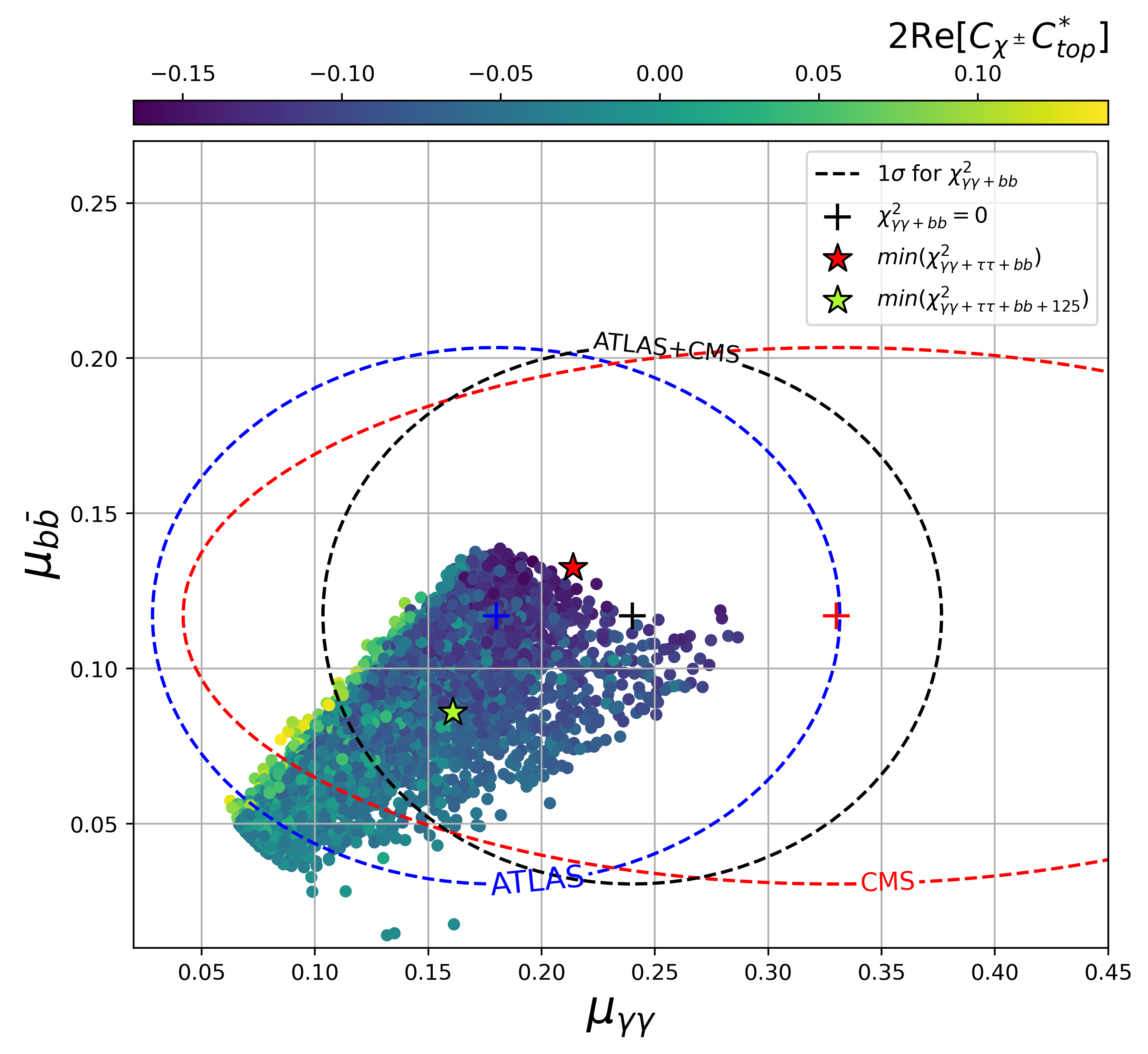}
\caption{Upper panel: squared moduli $|C_{H^\pm}|^2$ and $|C_{\chi^\pm}|^2$ as a function of  $\sin(\beta-\alpha)$, with the colour code indicating  $m_{H^\pm}$ and $m_{\chi^\pm}$. Lower panel: interference terms $2 \, \Re(C_{\chi^\pm} C^*_{H^\pm})$, $2 \, \Re(C_{\chi^\pm} C^*_{W^\pm})$, and $2 \, \Re(C_{\chi^\pm} C^*_{t})$ projected onto the $(\mu_{\gamma\gamma},\,\mu_{bb})$ plane. The modified top and $W^\pm$ contributions are also shown for comparison.}
\label{fig2}
\end{figure}

Clearly, the former contribution depends on the scalar mass $m_{\chi^\pm}$ and  reduced trilinear coupling  $h\chi^+\chi^-$.  This latter reads as 
\begin{equation}
\label{eq:h1coup_to_chipm}
\tilde{c}_{h\chi^+\chi^-}=-\frac{1}{g} \frac{m_W}{m_{\chi^\pm}^2} c_{h_1\chi^+\chi^-} \quad \text{where}\quad c_{h\chi^+\chi^-} = -\rho_a v \sin(\beta-\alpha),
\end{equation}
according to which substantial constructive interference between the $\chi^\pm$ and  $H^\pm$ loops  may occur, for non-vanishing $\sin(\beta-\alpha)$.

To highlight the effect of the inert charged scalars $\chi^\pm$ on the diphoton decay amplitude, and consequently on the enhancement of the $\mu_{\gamma\gamma}$ signal strength, we present in the upper panels of Fig.~\ref{fig2} the squared moduli $|C_{H^\pm}|^2$ and $|C_{\chi^\pm}|^2$ as a function of $\sin(\beta-\alpha)$, with the colour code indicating $m_{H^\pm}$ and $m_{\chi^\pm}$, respectively. For comparaison, the dominant SM contributions, i.e. $|C_{W}|^2$ and $|C_{top}|^2$, are also shown. It is evident that the $\chi^\pm$ loop contribution is, overall, significantly suppressed compared to that of $H^\pm$. However, for light $\chi^\pm$ masses, the corresponding loop contribution, can still interfere constructively with the dominant $H^\pm$ loop. This can lead to a non-negligible relative shift in $\mu_{\gamma\gamma}$ of a few percent in the region where $\sin(\beta - \alpha)$ slightly deviates from the alignment limit.

\begin{figure}[!h]
\centering
\begin{minipage}{0.38\textwidth}
\centering
\includegraphics[width=1.1\textwidth]{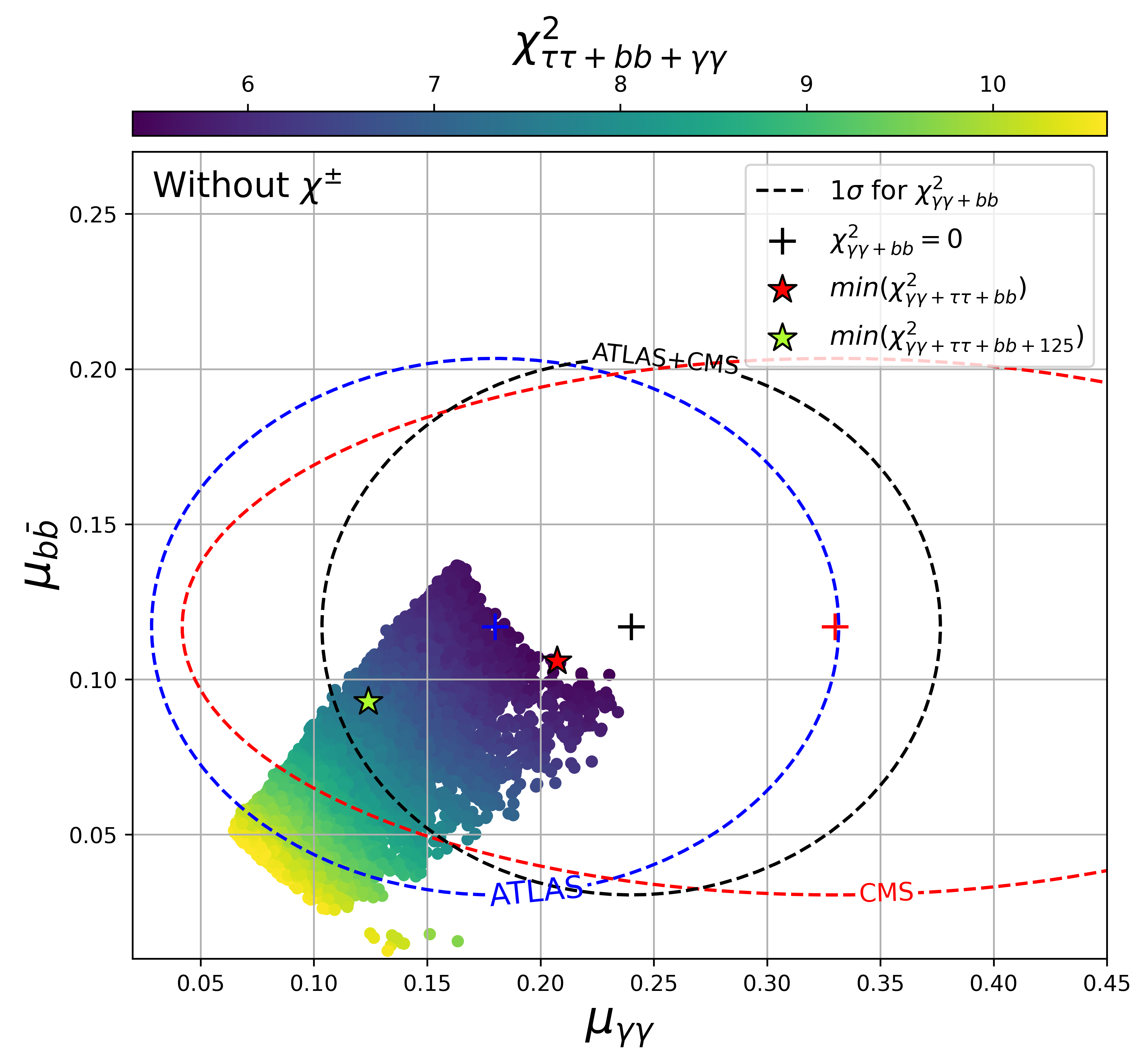}
\end{minipage}
\hspace*{1cm}
\begin{minipage}{0.4\textwidth}
\centering
\includegraphics[width=1.1\textwidth]{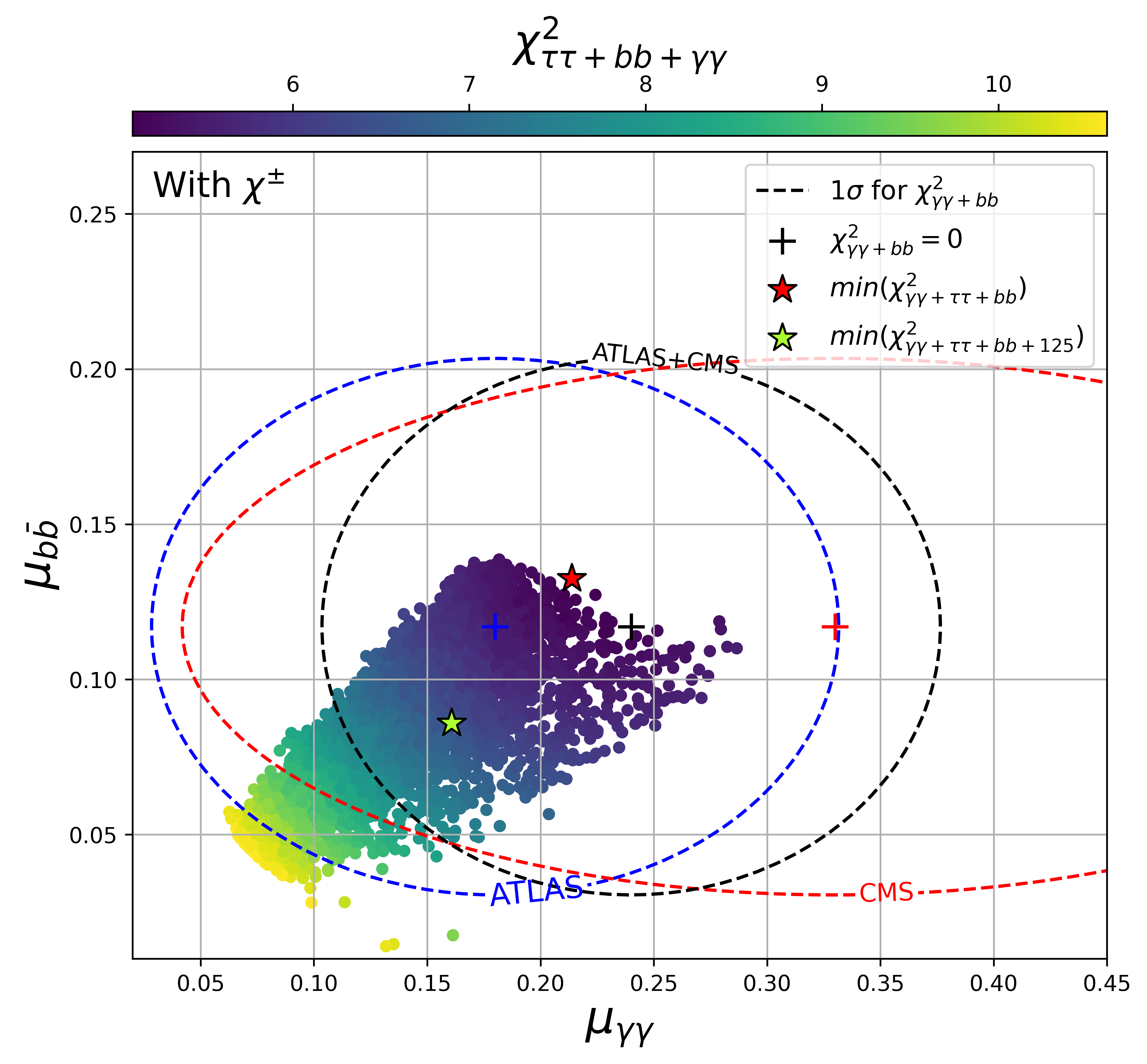}
\end{minipage}
\caption{Comparison of $\mu_{\gamma\gamma}-\mu_{bb}$ correlation with and without the $\chi^\pm$ contribution.}
\label{fig3}
\end{figure}

For the same purpose, the lower panels of Fig.~\ref{fig2} exhibit the projection of the interference terms $2 \, \Re(C_{\chi^\pm} C^*_{H^\pm})$, $2 \, \Re(C_{\chi^\pm} C^*_{W^\pm})$  and $2 \, \Re(C_{\chi^\pm} C^*_{t})$ within the $(\mu_{\gamma\gamma},\,\mu_{bb})$ plane. These plots make it clear that the interference between $\chi^\pm$ and $W^\pm$ is generally strong and varies substantially across the allowed parameter space. A mild but positive interference between $\chi^\pm$ and $H^\pm$ is also observed, which can enhance somewhat the $\mu_{\gamma\gamma}$ rate without conflicting with current Higgs precision measurements. Finally, the interference between $\chi^\pm$ and the top-quark loop remains relatively small and is typically destructive in the region that best fits the observed $\mu_{\gamma\gamma} $ excess.
In the light of the foregoing it is worth mentioning that, altogether, the $\chi^\pm$ contribution, even if small, has a non-negligible phenomenological impact in the I(1+2)HDM and is sufficiently relevant to justify going from the 2HDM to the I(1+2)HDM, given that its inclusion leads to a visible shift of the $\mu_{\gamma\gamma}$ and $\mu_{bb}$ distributions, as can be seen from Fig.~\ref{fig3}.

\begin{figure}[!ht]
\centering
\includegraphics[width=0.325\textwidth]{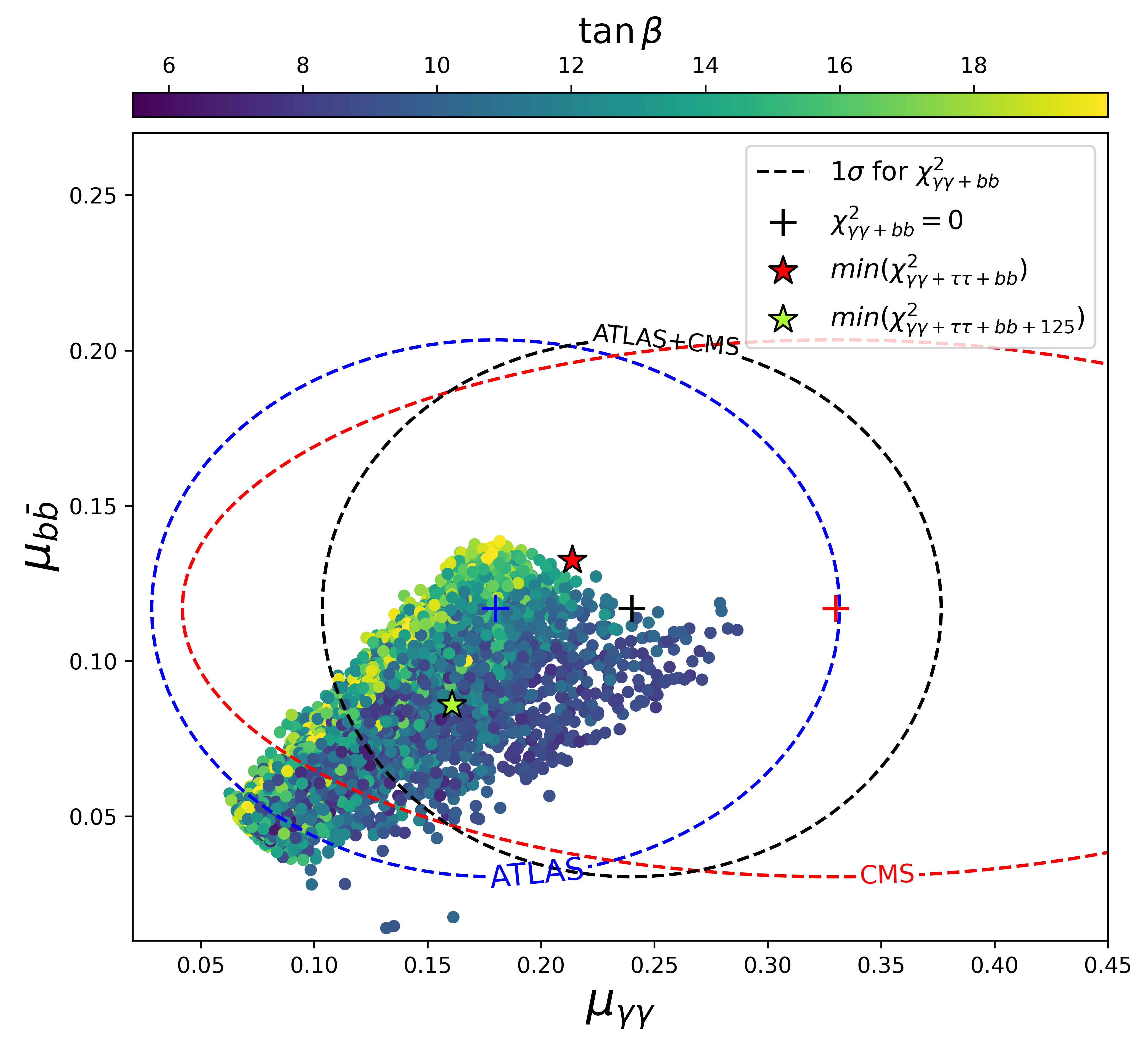}
\includegraphics[width=0.325\textwidth]{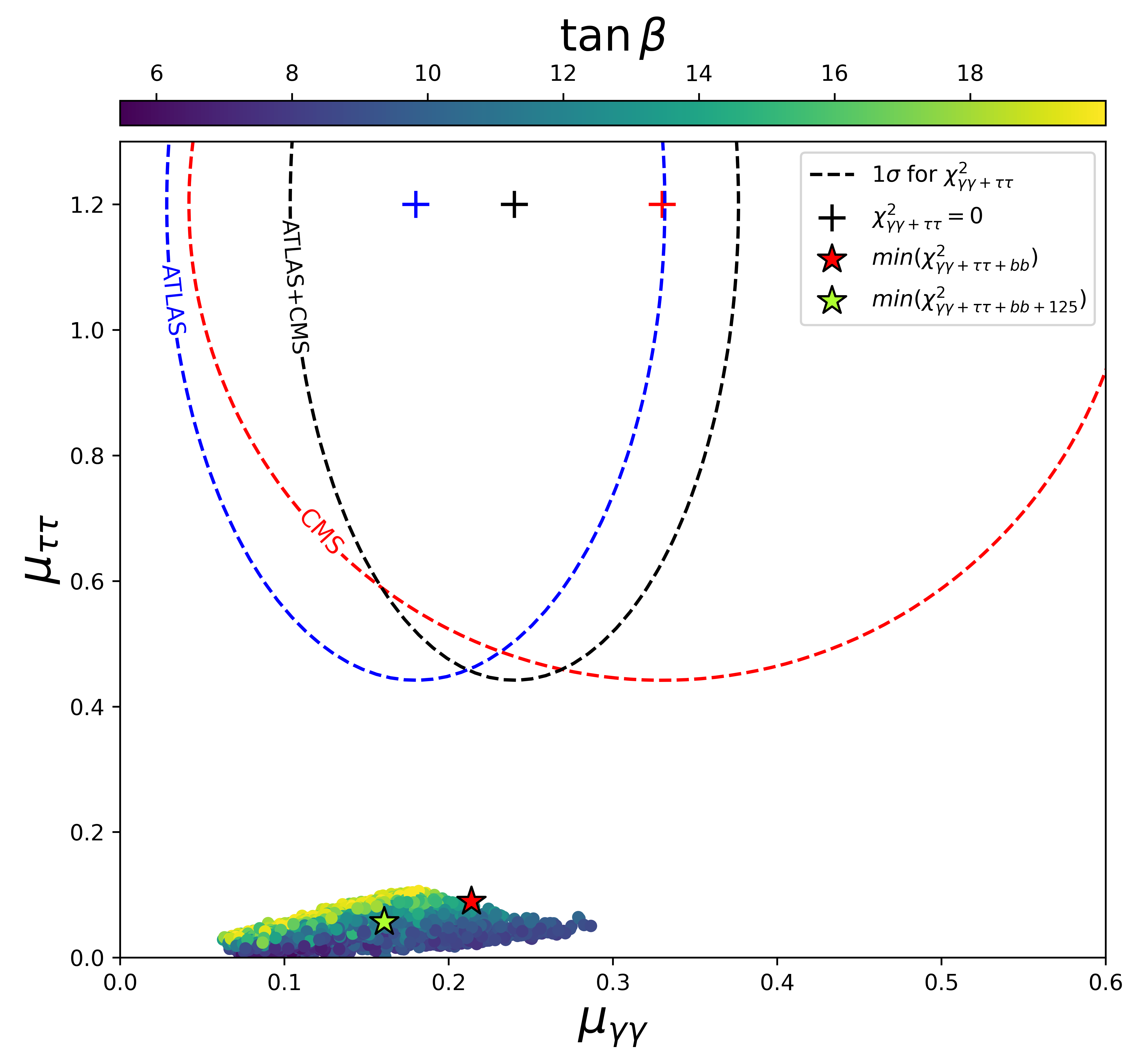}
\includegraphics[width=0.325\textwidth]{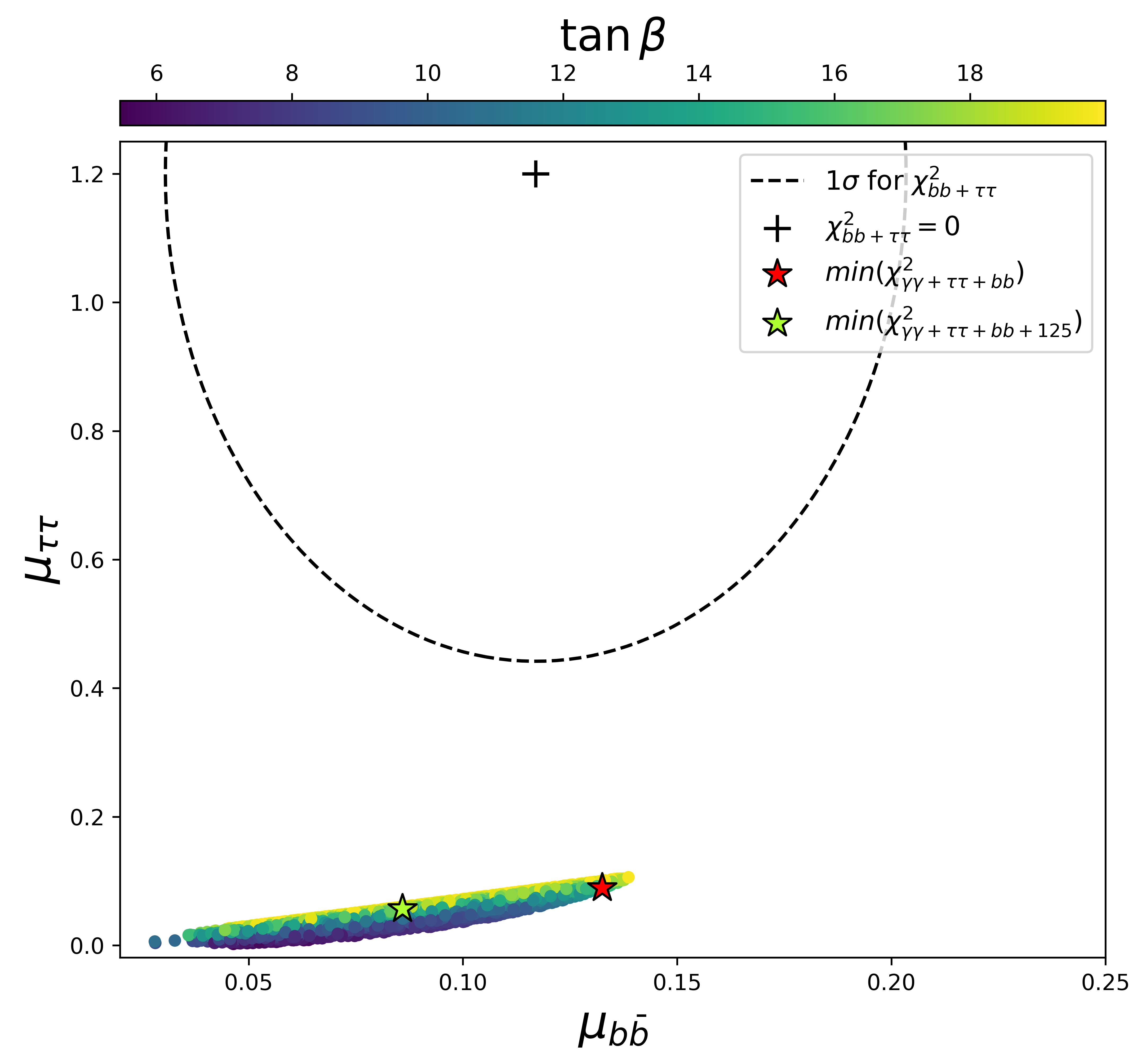}
\caption{Correlations among the signal strengths ($\mu_{bb}$ - $\mu_{\gamma\gamma}$), ($\mu_{\tau^+\tau^-}$ - $\mu_{\gamma\gamma}$) and ($\mu_{\tau^+\tau^-}$ - $\mu_{bb}$), with the $\tan\beta$ value represented by the colour bar . The ATLAS and CMS results, as described in the previous figure, are also included.} 
 \label{fig4}
\end{figure}

Another phenomenologically interesting quantity to study is $\tan\beta$, which influences nearly all the Higgs couplings. So, by considering the ATLAS and CMS results, along with their corresponding 1$\sigma$ uncertainties, we investigate the constraining power of each correlation on $\tan\beta$, and such behaviour is exhibited in Fig.~\ref{fig4}. So, a striking observation from the left plot is that many points fall within the 1$\sigma$ ellipse, indicating that both the $\mu_{bb}$ and $\mu_{\gamma\gamma}$ excesses are simultaneously enhanced. This behavior clearly tracks a controlled evolution with $\tan\beta$, allowing for a broad range, namely $5.3 \lesssim \tan\beta \lesssim 10.6$. This is in turn supported by the position, within the 1$\sigma$ contour, of the best-fit point that minimise $\chi^2_{\gamma\gamma+\tau^+\tau^-+b\bar{b}+125}$. On the other hand, the middle and right panels show that $\mu_{\tau^+\tau^-}$ tends to be smaller than the observed central value, $\mu_{\tau^+\tau^-}^{exp}=1.2$, throughout the explored range of $\tan\beta$. Nevertheless, higher values for such anomaly, even far from the experimental measurement can be achieved at larger $\tan\beta$, for a fixed $\mu_{\gamma\gamma}$. Altogether, combining all excesses, it may be inferred that $\tan\beta$ acts as a crucial adjustable parameter, providing valuable insights into the viable parameter space configurations of the I(1+2)HDM Type-I.
\begin{figure}[!h]	
\centering  
\includegraphics[width=0.45\textwidth]{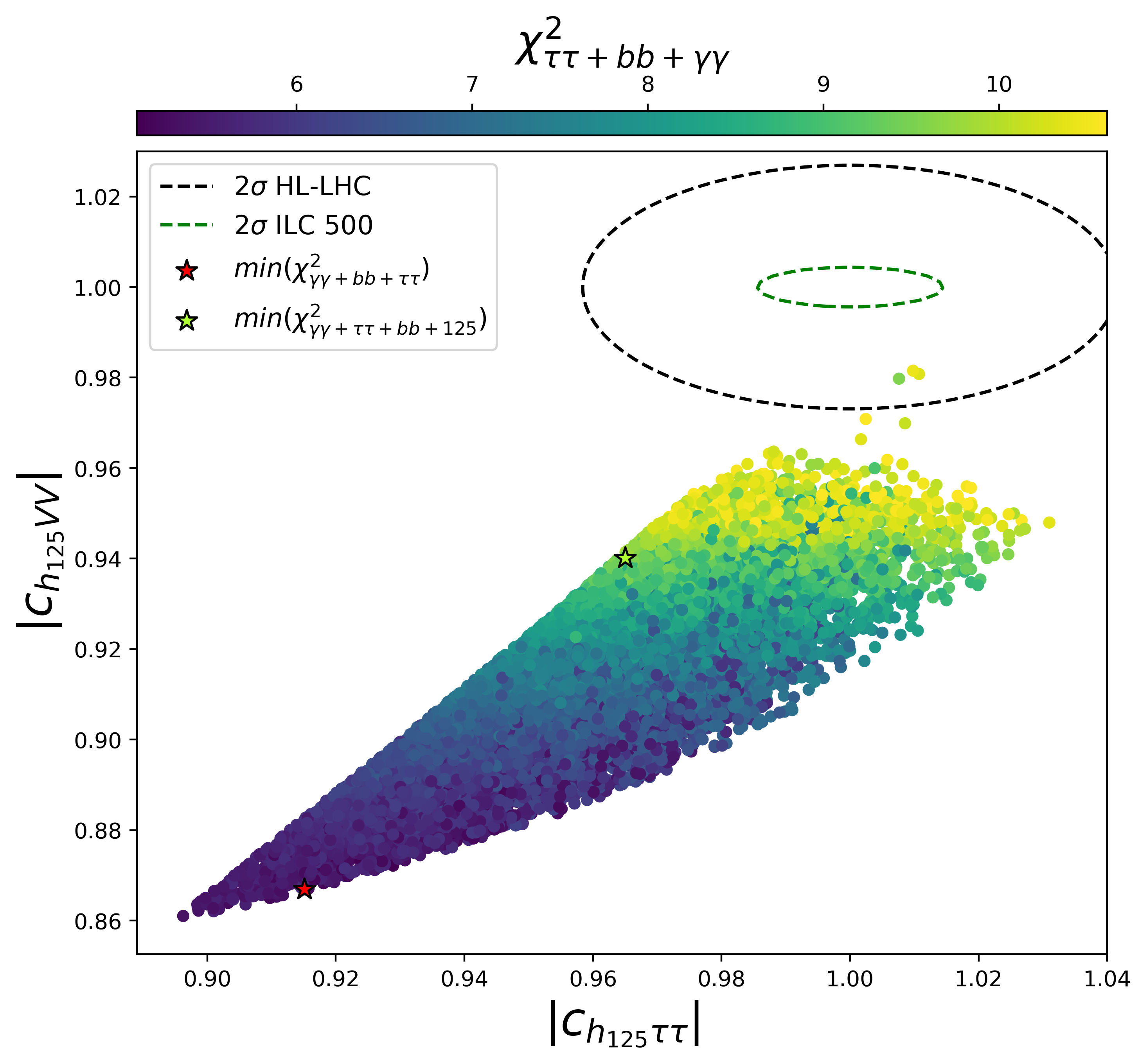}
\caption{Scattered points in the plane ($|c_{h_{125}\tau^+\tau^-}|$, $|c_{h_{125}VV}|$) following our scan, with the colour bar indicating the combined value of $\chi^2$ for the three discussed excesses. Projections from the HL-LHC and an ILC running at 500 GeV are also sketched.}
\label{fig5}
\end{figure}

Since the coupling of the SM-like Higgs boson to gauge bosons is extremely constraining for BSM scenarios (through mixing), the measurement of this quantity could afford one with some understanding of the observed excesses
at 95 GeV. In this connection, we display in Fig.~\ref{fig5} the $\chi^2_{\gamma\gamma+\tau^+\tau^-+b\bar{b}}$ distribution obtained from our parameter scan as a function of $|c_{h^{\rm SM}_{125}\tau^+\tau^-}|$ and $|c_{h^{\rm SM}_{125}VV}|$, the values of the $h^{\rm SM}_{125}\tau^+\tau^-$ and $h^{\rm SM}_{125}VV$ ($V=W^\pm, Z$) couplings (in modulus), respectively,  in the I(1+2)HDM Type-I. The magenta and green ellipses indicate the projected uncertainties achievable at the High-Luminosity LHC (HL-LHC) \cite{Gianotti:2002xx,Cepeda:2019klc} (with 3 ab$^{-1}$ of integrated luminosity) and a possible International Linear Collider
(ILC) running at a center-of-mass energy of 500 GeV \cite{Bambade:2019fyw} (with 1 ab$^{-1}$ of integrated luminosity). The  SM prediction for both coupling coefficients is marked by a ``+". 

The data sample here fulfills both {{ $\chi^2_{\gamma\gamma+\tau^+\tau^-+b\bar{b}} \le 8.03$ and $\chi^2_{125}-min(\chi^2_{125}) \le 6.18$}}, with the results from the I(1+2)HDM Type-I showing that most points are concentrated at low values, well below unity for both $|c_{h_{125}\tau^+\tau^-}|$  and  $|c_{h_{125}VV}|$. This is related to the fact that the minimum of $\chi^2_{\gamma\gamma+\tau^+\tau^-+b\bar{b}}$ (red star) requires a non-vanishing value of $\sin(\beta-\alpha)$ (so that the $\chi^\pm$ contribution cooperates with the $H^\pm$ one in $h\to\gamma\gamma$, as previously explained), hence,  $|c_{h_{125}VV}|=|\cos(\beta-\alpha)|<1$. Only a tiny portion of this BSM parameter space is potentially compatible with 
HL-LHC and ILC (projected) measurements (assuming the same central values of $|c_{h_{125}\tau^+\tau^-}|$  and  $|c_{h_{125}VV}|$ as presently), so that either of these machines will prove a crucial testing ground of the viability of our model (assuming persistence of the current $\gamma\gamma$, $b\bar b$ and $\tau^+\tau^-$ excesses at the current level by the end of the LHC runs).

\begin{figure}[!h]	
\centering
\includegraphics[width=0.4\textwidth]{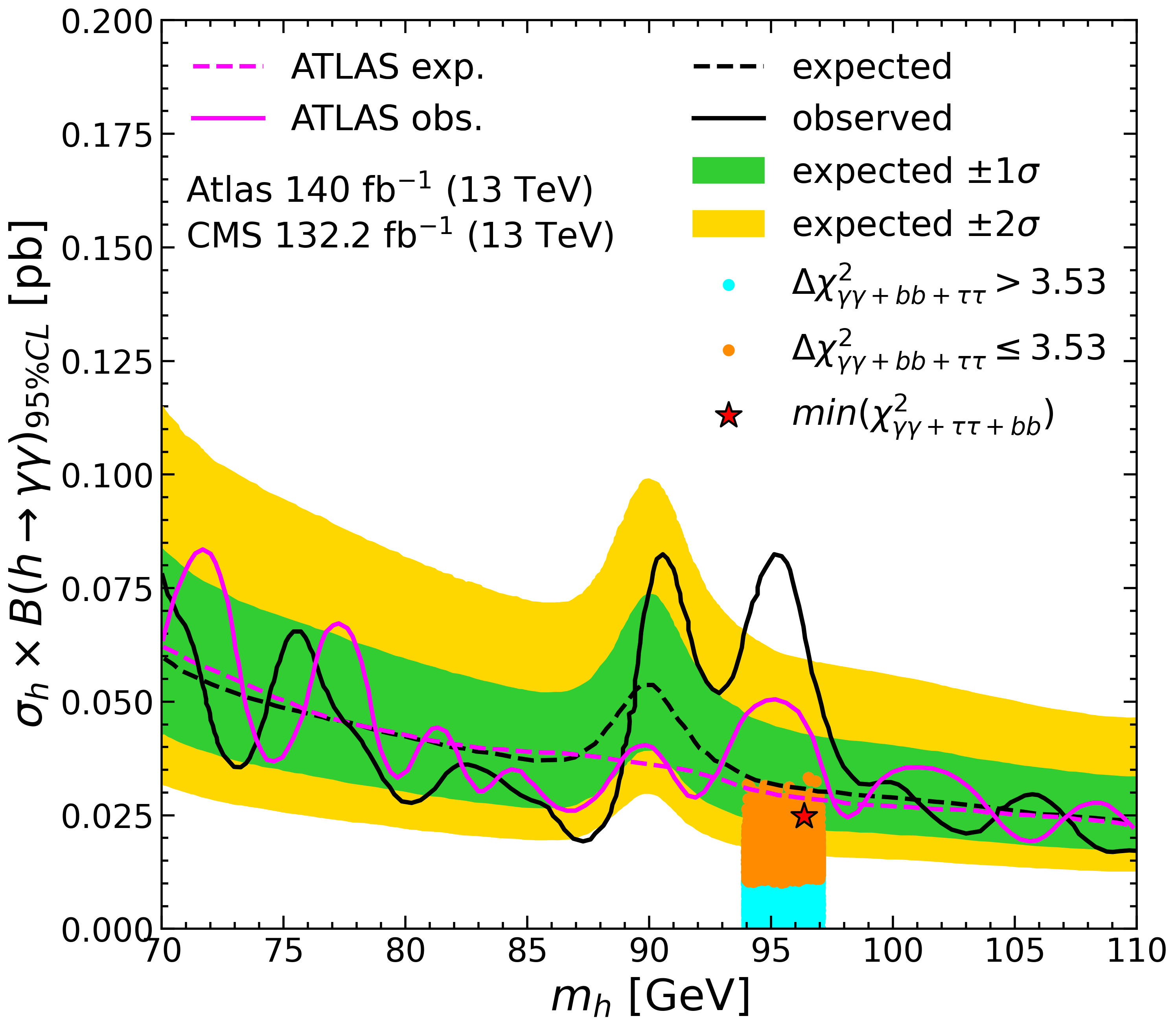}			
\caption{Scattered points in the $[m_{h}$, $\sigma_h \times {\rm BR}(h \to \gamma\gamma)]$ plane. The solid and dashed lines indicate the observed and expected cross section limits obtained by ATLAS and CMS, respectively. ATLAS is depicted in magenta, while CMS is shown in black. The green and yellow bands represent the $1\sigma$ and $2\sigma$ uncertainty intervals, respectively. The orange points feature $\Delta\chi^2_{\gamma\gamma + bb +\tau\tau} \leq 3.53$ and describe the excesses at the level of $1\,\sigma$ ($ 68.27 \% $ C.L.) or better, whereas the cyan points represent $\Delta\chi^2_{\gamma\gamma + bb +\tau\tau} > 3.53$.}
\label{fig6}
\end{figure}

Finally, as intimated, we dwell in some more detail into the $\gamma\gamma$ excess, 
by presenting in Fig.~\ref{fig6} a superposition of our allowed parameter points and experimental results, by overlaying the former on the combined ATLAS and CMS low mass $\gamma\gamma$ analysis data at 13 TeV, with integrated luminosities of 140 fb$^{-1}$ and 132.2 fb$^{-1}$ \cite{CMS:2023yay, Arcangeletti:2023}. The expected and observed ATLAS limits on the cross section times Branching Ratio (BR), i.e., $\sigma(pp \to h)\times {\rm BR} (h \to \gamma\gamma)$, are indicated by the magenta dashed and solid lines, respectively. The 1$\sigma$ and 2$\sigma$ uncertainty bands, resulting from the combined result, are shown in green and yellow, respectively. Superposed on these results, the combined (between ATLAS and CMS) expected and observed limits are illustrated by the black dashed and solid lines, respectively. Altogether, the plot clearly indicates that our parameter points are not only in very good agreement with the LHC excess in the di-photon channel but also that our sample matches well all current anomalous data, as emphasised by the position of the best-fit point explaining both the $\gamma\gamma$ and $b\bar{b}$ excesses simultaneously. To summarise the previous results, the details of our best-fit point are presented in Tab.~\ref{description_bestfit}.
\begin{table*}[!h] 
{\footnotesize	
\setlength{\tabcolsep}{0.1cm}
\renewcommand*{\arraystretch}{1.8}
\begin{tabular*}{\textwidth}{ @{\extracolsep{\fill}} lccccccccccccccc}\noalign{\hrule height 0.8pt}
\noalign{\vspace{1.25pt}}  
\noalign{\hrule height 0.3pt}
\textbf{Parameters} &
$m_{h}$ & $m_{H}$ & $m_{A}$& $m_{H^{\pm}}$ & $\tan\beta$ & $\sin(\beta - \alpha)$ & $m_{12}^2$ & $m_{\chi}$ &  $m_{\chi_{a}}$ & $m_{\chi^{\pm}}$ & $\rho_{\eta}$ & $m_{\eta}^2$ & \\
\noalign{\hrule height 0.9pt}
&96.35 & 125.09 & 241.26 & 166.98 & 14.29 &  -0.36 &  447.44 &  312.67 &  530.02 &  566.25 & 8.39 & $35990.77$  \\
\noalign{\vspace{1.25pt}}  \noalign{\hrule height 0.9pt}
\end{tabular*}
{\setlength{\tabcolsep}{0.1cm}
\begin{tabular*}{\textwidth}{@{\extracolsep{\fill}}lccccccccccccccccc}
\textbf{Signal strengths} &
$\mu_{\gamma\gamma}$ &&$\mu_{\tau^+\tau^-}$&&$\mu_{b\bar{b}}$\\
\noalign{\hrule height 0.9pt}
& 0.21 &&  0.09    &&       0.13\\
\noalign{\vspace{1.25pt}}  \noalign{\hrule height 0.9pt}	
\end{tabular*}}
\caption{\small  Details of the best-fit point corresponding to $\min(\chi^2_{\gamma\gamma+\tau^+\tau^-+b\bar{b}})=5.09$. (Mass (squared) are in GeV$^{(2)}$.) }
\label{description_bestfit}
}
\end{table*}
Furthermore, it is worth mentioning that the I(1+2)HDM Type-I is not only able to explain the current excesses but also has the potential to give distinctive signals at future accelerators,  such as an $e^+e^-$ collider. Our findings in Tab.~\ref{CS_bestfit} demonstrate that such a machine will have the capability to probe this new state through various production mechanisms, primarily Higgs-strahlung and Vector Boson Fusion (VBF) processes. Indeed, at $\sqrt{s}=250$ GeV, the $e^+e^- \to Z^{\ast} \to Z h_{95}$ process dominates with a sizable cross section of 47.78 $fb$, making it the most promising channel for future discovery. As the center-of-mass-energy increases, the VBF channels become increasingly more relevant, with $\sigma(e^+e^- \to \nu_e \bar{\nu}_e h_{95})$ reaching 12.96 $fb$ at $\sqrt{s}=500$ GeV. Similarly, the $e^+e^- \to e^+e^-  h_{95}$ process exhibits a steady rise, reaching nearly 0.845 $fb$ at 500 GeV. These cross sections indicate that future $e^+e^-$ colliders would provide a powerful testing ground for the I(1+2)HDM Type-I, by enabling precise measurements of the properties of the observed 95 GeV scalar state.

\begin{table}[!h]
\centering
\begin{tabular}{lrrr}
\toprule 
$\sqrt{s}=$     & 250 GeV & 350 GeV & 500 GeV \\ \midrule
$\sigma(e^+ e^-\to Z h_{95})$                & 47.78 fb         & 20.07 fb           & 8.24 fb         \\
$\sigma(e^+ e^-\to \nu_{e} \bar{\nu_{e}} h_{95})$        & 2.27 fb          & 6.29 fb         & 12.96 fb       \\
$\sigma(e^+ e^-\to e^+ e^- h_{95})$              & 0.182 fb           & 0.478 fb          & 0.845 fb        \\ \midrule
\end{tabular}
\caption{\small  Cross sections for  Higgs-strahlung, $W^+W^-$- and Z-fusion
processes for $h_{95}$ at the three centre-of-mass energies corresponding to the best-fit point $\min(\chi^2_{\gamma\gamma+\tau^+\tau^-+b\bar{b}})=5.09$.}
\label{CS_bestfit}
\end{table}

\section{Conclusions}
\label{sec:conclusion}

In summary, we have looked at how to address the $\gamma\gamma$, $b\bar{b}$ and $\tau^+\tau^-$ excesses observed around 95 GeV by the LHC and LEP, within the I(1+2)HDM Type-I, while remaining consistent with measured properties of the Higgs boson observed at 125 GeV as well as other theoretical and experimental constraints. Rigorous data evaluations, complemented by detailed simulations and advanced computational methods, have been carried out, followed by a study of the correlations between the signal strengths  $\mu_{\gamma\gamma}$, $\mu_{bb}$ and $\mu_{\tau^+\tau^-}$ at the $1\sigma$ C.L. 

Our findings suggest that all two-dimensional correlations, i.e, on the planes ($\mu_{x}$-$\mu_{y}$), with $x,y=\gamma\gamma,\,b\bar{b},\tau^+\tau^-$, incorporate many points lying entirely within the 1$\sigma$ C.L., indicating that our model can simultaneously explain both the $\gamma\gamma$ and $b\bar{b}$ excesses, particularly for low values of $\tan\beta$. Besides, our study indicates there is a crucial distinction compared to results found in \cite{Khanna:2024bah}, wherein the 2HDM Type-I was found to struggle in accommodating all three excesses simultaneously within a viable parameter space so that embedding an extra inert doublet significantly enhances the theory compatibility with data.

However, as the LHC has not yet completed its Run 3, extended data collection will provide a valuable opportunity to further investigate the excesses observed around 95 GeV and also to clarify the interplay between the coupling measurements of the $h_{95}^{\rm NP}$ (the $h$ state of our BSM scenario) and $h_{125}^{\rm SM}$ (the $H$ state of our BSM scenario), which will be crucial in assertion or not of such new physics. However, we have also demonstrated that, if the level of confidence of these three excesses remains so at the end of the LHC era, both the HL-LHC and ILC will be able to finally access the viability of the I(1+2)HDM Type-I as the underlying BSM scenario, by either exploiting further the
excesses associated to the $h_{95}^{\rm NP}$  state or else by ever more precise measurements of the $h_{125}^{\rm SM}$ state.

Finally, we have shown that any future $e^+e^-$ collider running at approximately 250, 350 and 500 GeV will have a strong sensitivity to the best-fit point of our model to the current anomalies at 95 GeV, by exploiting the direct production of the $h_{95}^{\rm NP}$ state in both Higgs-strahlung and VBF.

\section*{Acknowledgments}
SM is supported in part through the NExT Institute and  STFC Consolidated Grant ST/X000583/1. AH and LR would like to thank EL-said Ghourmin for invaluable discussions.

\bibliography{refs} 
\bibliographystyle{JHEP}

\end{document}